\documentclass[lettersize,journal]{IEEEtran}

\usepackage{amsmath,amsfonts}
\usepackage{algorithmic}
\usepackage{algorithm}
\usepackage{array}
\usepackage[caption=false,font=normalsize,labelfont=sf,textfont=sf]{subfig}
\usepackage{textcomp}
\usepackage{stfloats}
\usepackage{url}
\usepackage{verbatim}
\usepackage{graphicx}
\usepackage{cite}
\usepackage[printonlyused]{acronym}
\newacro{los} [LOS] {line-of-sight}
\newacro{em} [EM] {electromagnetics}
\newacro{ml} [ML] {maximum likelihood}
\newacro{kpi} [KPI] {key performance indicator}
\newacro{mse} [MSE] {mean square error}
\newacro{aod} [AoD] {angle-of-departure}
\newacro{dof} [DoF] {degree-of-freedom}
\newacro{upa} [UPA] {uniform planar array}
\newacro{snr} [SNR] {signal-to-noise ratio}
\newacro{rmse} [RMSE] {root mean square error}
\newacro{crlb} [CRLB] {Cram\'er-Rao lower bound}
\newacro{kld} [KLD] {Kullback–Leibler divergence}
\newacro{siso} [SISO] {single-input-single-output}
\newacro{mimo} [MIMO] {multiple-input multiple-output}
\newacro{mcrb} [MCRB] {misspecified Cram\'er-Rao bound}
\newacro{ris} [RIS] {reconfigurable intelligent surface}
\usepackage{color} 
\newcommand{\red}[1]{{\color{red}{#1}}} 

\usepackage{tikz} 
\usepackage[utf8]{inputenc}
\usepackage{pgfplots} 
\makeatletter
\newcommand{\gettikzxy}[3]{%
  \tikz@scan@one@point\pgfutil@firstofone#1\relax
  \edef#2{\the\pgf@x}%
  \edef#3{\the\pgf@y}%
}
\usepackage{bchart}
\usepackage{tabu,longtable}
\usepackage{diagbox}
\usepackage{threeparttable}
\usepackage{makecell}
\usepackage{multirow}

\usepackage{units} 
\usepackage{bm} 
\usepackage{amssymb} 
\newcommand{\TT}{\mathsf{T}}
\newcommand{\HH}{\mathsf{H}}

\newcommand{\av}{{\bf a}}

\newcommand{\ev}{{\bf e}}

\newcommand{\gv}{{\bf g}}
\newcommand{\hv}{{\bf h}}

\newcommand{\pv}{{\bf p}}

\newcommand{\rv}{{\bf r}}
\newcommand{\sv}{{\bf s}}

\newcommand{\uv}{{\bf u}}

\newcommand{\vv}{{\bf v}}
\newcommand{\xv}{{\bf x}}
\newcommand{\yv}{{\bf y}}

\newcommand{\Am}{{\bf A}}

\newcommand{\Hm}{{\bf H}}

\newcommand{\Qm}{{\bf Q}}

\newcommand{\Sm}{{\bf S}}

\newcommand{\Et}{{\rm E}}
\newcommand{\Ft}{{\rm F}}

\newcommand{\Pt}{{\rm P}}

\newcommand{\Rt}{{\rm R}}
\newcommand{\St}{{\rm S}}

\newcommand{\thetav}{\hbox{$\boldsymbol\theta$}}

\newcommand{\Thetam}{\hbox{\boldmath$\Theta$}}

\begin{document}
\title{Mutual Coupling in RIS-Aided Communication: Model Training and Experimental Validation}

\author{Pinjun~Zheng, 
Ruiqi~Wang, 
Atif~Shamim,~\IEEEmembership{Fellow,~IEEE}, 
and~Tareq~Y.~Al-Naffouri,~\IEEEmembership{Senior Member,~IEEE}

\thanks{The authors are with the Electrical and Computer Engineering Program, Division of Computer, Electrical and Mathematical Sciences and Engineering (CEMSE), King Abdullah University of Science and Technology (KAUST), Thuwal, 23955-6900, Kingdom of Saudi Arabia.}
}

\markboth{draft}{draft}
\maketitle

\begin{abstract}
Mutual coupling is increasingly important in reconfigurable intelligent surface (RIS)-aided communications, particularly when RIS elements are densely integrated in applications such as holographic communications. This paper experimentally investigates the mutual coupling effect among RIS elements using a mutual coupling-aware communication model based on scattering matrices. Utilizing a fabricated 1-bit quasi-passive RIS prototype operating in the mmWave band, we propose a practical model training approach based on a single 3D full-wave simulation of the RIS radiation pattern, which enables the estimation of the scattering matrix among RIS unit cells. The formulated estimation problem is rigorously convex with a limited number of unknowns un-scaling with RIS size. The trained model is validated through both full-wave simulations and experimental measurements on the fabricated RIS prototype. Compared to the conventional communication model that does not account for mutual coupling in RIS, the mutual coupling-aware model incorporating trained scattering parameters demonstrates improved prediction accuracy. Benchmarked against the full-wave simulated RIS radiation pattern, the trained model can reduce prediction error by up to approximately 10.7\%. Meanwhile, the S-parameter between the Tx and Rx antennas is measured, validating that the trained model exhibits closer alignment with the experimental measurements. These results affirm the accuracy of the adopted model and the effectiveness of the proposed model training method.
\end{abstract}


\section{Introduction}

Standing as a groundbreaking technique in modern wireless communication systems, \acp{ris} represent a revolutionary paradigm for the smart \ac{em} wave control~\cite{Di2022Communication,Pan2022An,Pei2021RIS,Cheng2023Degree}. These surfaces, composed of an array of programmable reflective or refractive unit cells, empower real-time adjustments to the radio environment. Leveraging advanced metamaterials and signal processing techniques, RIS enables precise modulation of the phase, amplitude, and polarization of \ac{em} waves. Such adaptability promises to significantly benefit wireless applications including communication, sensing, localization, and mapping~\cite{Bjornson2022Reconfigurable,Song2023Intelligent,Zheng2023JrCUP,Chen2023Multi,Zheng2023LEO}. With the advent of densely integrated RISs such as holographic RIS~\cite{Wan2021Terahertz}, the phenomenon of mutual coupling emerges as a critical consideration with profound implications for the performance of RIS-assisted communications~\cite{Di2022Communication}. Mutual coupling refers to the interaction between adjacent RIS unit cells, wherein the \ac{em} fields generated by one unit cell influence the behavior of its neighboring unit cells. This inter-element coupling effect can significantly impact the overall functionality and effectiveness of a RIS deployment, especially in scenarios involving high amplitude gains or varying operational frequencies. Therefore, understanding and mitigating the impact of mutual coupling is essential for optimizing the performance of RIS systems. 

\subsection{Prior Work and Motivation}

RIS-aided communication has been extensively studied in recent years owing to its substantial potential in extending communication coverage~\cite{Sang2022Coverage}, enhancing signal reception~\cite{Zhang2023Active}, and improving energy efficiency~\cite{Huang2019Reconfigurable}. 
To date, numerous novel signal processing techniques have been developed to support RIS-aided wireless applications~\cite{Pan2022An,He2021Channel,Zheng2023JrCUP}. 
	 {Besides those efforts in signal processing techniques, significant advancements have been made in \ac{ris} hardware. On the one hand, some designs focus on novel RIS concepts and implementations. For instance, a wideband RIS that fully covers the n257 and n258 5G mmWave bands was reported in~\cite{Wang2023Wideband}. The authors of~\cite{Rao2023An} designed an active RIS with consistent reflection polarization using reflection amplifiers. Furthermore, the authors of~\cite{Hu2024Methodology} proposed a RIS design featuring absorptive filtering capabilities to suppress interference. On the other hand, some RIS developments incorporate advanced fabrication and control techniques. For example, the authors of~\cite{Sayanskiy2023A} designed an RIS with robust remote infrared control. Additionally, the authors of~\cite{Yang2023A} and~\cite{Yang2024Fully} introduced the concept of fully printed RIS, including both metallic patterns and switches, which substantially reduces fabrication costs, especially for large-scale RISs.}


 {Regarding the RIS mutual coupling, numerous recent investigations  have led to significant advancements in modeling and performance optimization. A physics-based end-to-end model, named PhysFad, has been proposed in~\cite{Faqiri2023PhysFad} based on a first-principles coupled-dipole formalism. This model characterizes RIS-parametrized wireless channels with adjustable fading. Additionally, a linear series expansion-based analysis of the channel's non-linearity mechanism has been presented in~\cite{Rabault2024On}, and an experimental validation using a 1-bit RIS prototype operating in the cmWave band has been reported in~\cite{Sol2024Experimentally}. Stemming from fundamental \ac{em} principles, the PhysFad model inherently accounts for proximity-induced mutual coupling, reverberation-induced multi-reflection, and environmental fading, providing essential insights into accurate \ac{ris} channel modeling, particularly in rich-scattering environments. With its EM-compliant characteristics, this channel model serves as a reliable benchmark for generating physically precise wireless channels and verifying advanced RIS configuration designs~\cite{Stylianopoulos2022Deep}.}

Another suitable theory to model mutual coupling in RIS is the microwave network theory. Notably, existing microwave network-based communication modeling can be categorized into two types: impedance matrix-based models and scattering matrix-based models. In the impedance matrix-based approach, the first effort can be found in~\cite{Gradoni2021End}, where an \ac{em}-compliant and mutual coupling-aware communication model is proposed by adopting the mutual impedance analysis among RIS unit cells. 
This model has proven effective in guiding RIS configuration, facilitating applications such as end-to-end received power maximization in \ac{siso} systems~\cite{Qian2021Mutual} and sum-rate optimization in multi-user interference \ac{mimo} channels~\cite{Abrardo2021MIMO}. On the other hand, the scattering matrix-based model was initially explored by the authors of~\cite{Shen2022Modeling} through a scattering parameter network analysis. Subsequently, the authors of~\cite{Li2023Beyond2} derived a scattering matrix-based mutual coupling-aware communication model.  {The scattering parameter is regarded as a model that is more directly related to the radiation pattern and easier to measure~\cite{Wijekoon2024Phase}.} Importantly, the derivation in~\cite{Li2023Beyond2}  {and~\cite{Abrardo2023Design}} establish the equivalence between the impedance matrix-based and scattering matrix-based communication models.  {Recently, microwave network-based models, either in the impedance matrix form or the scattering matrix form, have been successfully applied, for example, in \ac{em} modeling for arbitrarily shaped \ac{ris}\cite{Badheka2023Accurate} and in \ac{ris}-parametrized wireless networks-on-chip~\cite{Tapie2024Systematic}.}

 {This paper aims to provide a practical method for evaluating mutual coupling in RIS, thereby facilitating channel modeling and downstream applications. Due to the passive reflective nature of RIS, the accurate coupling parameters between its unit cells cannot be directly acquired through simulation or measurement. 
Notice that in far-field and sparse-scattering environments, the wireless channel can be accurately characterized by the conventional cascaded channel model, as evidenced by various studies~\cite{Nayeri2018Reflectarray,Tang2020MIMO,Brady2013Beamspace}. In pursuit of both accurate mutual coupling description and model tractability, this paper adopts microwave network theory-based modeling to propose a novel scattering parameter estimation method. This method features a limited number of unknowns, independent of the number of \ac{ris} elements, as long as it constitutes a uniform array. Tailored for sparse-scattering propagation in mmWave/THz frequency, the adopted microwave network theory enables the separation of RIS mutual coupling and environmental scattering in a cascaded manner, with negligible loss of accuracy, by disregarding mutual coupling with the transmitter and receiver under the far-field assumption~\cite{Abrardo2023Design, Shen2022Modeling}. Leveraging this model, the proposed approach can determine these limited scattering parameters based on the sample observations from a single full-wave simulation. The accuracy of the estimated results is validated through both simulation and experiment, encompassing assessment of the cross-angle radiation pattern and end-to-end S-parameter predictions.}


\subsection{Main Contributions}
The main contributions of this paper are summarized as follows.

\textit{Model Training:} Utilizing the microwave network-based channel model,
a realistic evaluation method of the coupling parameters among RIS unit cells is proposed. By only considering the coupling between adjacent RIS unit cells and neglecting very weak coupling effects beyond the distance of two unit cells, we parameterize the scattering matrix of the RIS into a finite set of scattering parameters. Based on this parameterized approximation, along with the far-field cascaded wireless channel model, we design a training approach to estimate these scattering parameters based on a single 3D full-wave \ac{em} simulation of the radiation pattern. The simulated RIS radiation pattern accurately captures the 3D \ac{em} behavior including mutual coupling, thus we can estimate these scattering parameters by optimizing their fitness to the simulated pattern. We show that the estimation can be formulated into a convex optimization problem via the Neumann series expansion, accompanied by a rigorous proof of convexity.  

\textit{Simulation Validation:} The mutual coupling-aware communication model with the trained scattering parameters is validated through the full-wave simulation. First, we perform the full-wave simulations of 3 different far-field radiation patterns of the RIS by varying the position of the Tx antenna. In the meantime, the theoretical predictions of these radiation patterns are obtained using the mutual coupling-aware model (with the trained scattering parameters) and the conventional mutual coupling-unaware model, respectively. Benchmarked against the simulated radiation patterns, the trained model exhibits superior accuracy compared to the conventional model without accounting for mutual coupling, validating the precision of the trained model.

\textit{Experimental Validation:} To further confirm the trained model, experimental validation is also carried out in this work. The experimental measurements are conducted on the 1-bit mmWave RIS prototype reported in~\cite{Wang2023Wideband}. We use two horn antennas to respectively transmit and receive \ac{em} signals, and measure the S-parameter between the two horn antennas through a vector network analyzer. Afterward, this S-parameter is predicted by the trained model and the conventional model, respectively. The comparisons over different setups demonstrate that the trained model can predict the S-parameter closer to the real measurement, revealing both the accuracy of the adopted mutual coupling-aware communication model and the effectiveness of the proposed scattering parameters training approach. Moreover, our experimental results demonstrate that while these S-parameters are derived from far-field modeling, they can also accurately predict near-field radiation.


The paper is organized as follows. Section II introduces the fundamentals of RIS hardware. Section III recaps the RIS-aided communication models, including both the conventional mutual coupling-unaware model and the recently proposed mutual coupling-aware model based on scattering matrices. Leveraging a 3D full-wave simulation of the RIS radiation pattern, a training approach for the scattering parameters among RIS unit cells is proposed in Section~IV. Based on the trained model, the simulation validation and the experimental validation are conducted in Section~V and Section~VI, respectively. Finally, the conclusions of the paper are drawn in Section VII.

\section{Fundamentals of RIS Hardware}

\subsection{Functional Principle of RIS}\label{sec_IIA}

\begin{figure}[t]
    \centering
    \begin{tikzpicture}
    \node (image) [anchor=south west]{\includegraphics[width=0.8\linewidth]{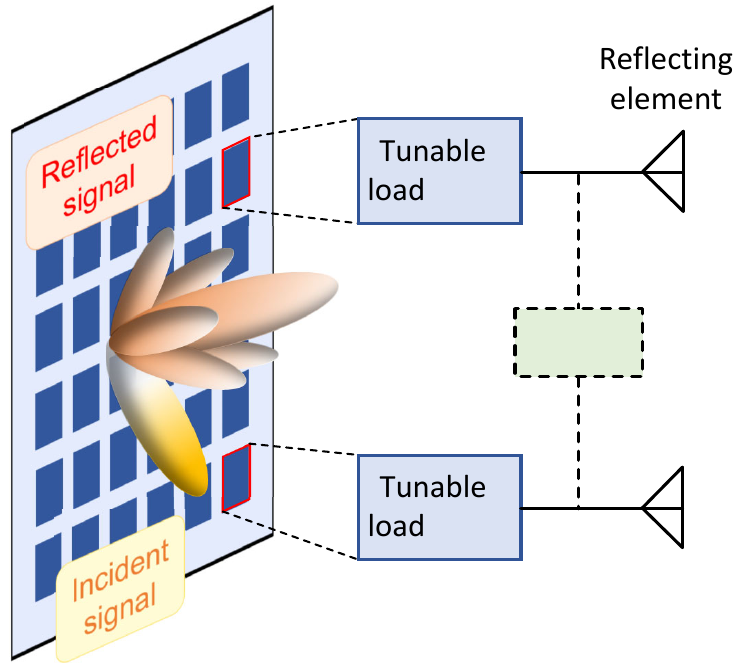}};
    \gettikzxy{(image.north east)}{\ix}{\iy};
    \node at (0.635*\ix,0.74*\iy)[rotate=0,anchor=north]{\footnotesize{$Z_{\text{RIS},m}$}};
    \node at (0.77*\ix,0.525*\iy)[rotate=0,anchor=north]{\footnotesize{$Z_{\text{RIS},m,n}$}};
    \node at (0.635*\ix,0.255*\iy)[rotate=0,anchor=north]{\footnotesize{$Z_{\text{RIS},n}$}};
    \node at (0.58*\ix,0.52*\iy)[rotate=0,anchor=north]{\large{$\dots$}};
	\end{tikzpicture}
    \caption{Conceptual architecture of a \ac{ris}.}
    \label{fig_RIS}
\end{figure}
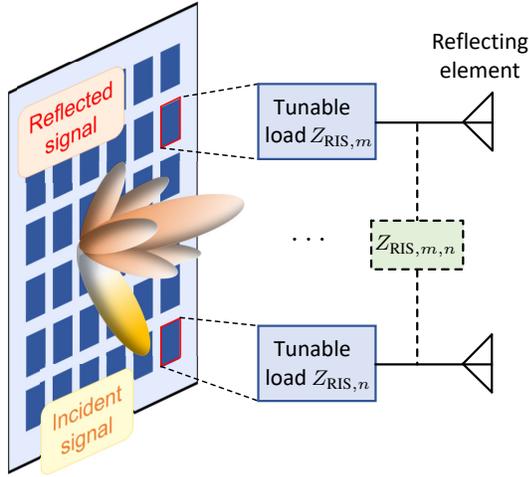

A \ac{ris} is a planar surface that consists of an array of scattering unit cells, each of which can impose an independent phase shift, and possibly an amplitude gain, on the incident \ac{em} waves. Fig.~\ref{fig_RIS} depicts a conceptual structure of the RIS. As illustrated, each RIS unit cell is equivalently comprised of a reflecting element and a functional circuit with tunable load $Z_{\text{RIS},m}$~\cite{Shen2022Modeling,Dai2020Reconfigurable,ElMossallamy2020Reconfigurable}, where $m$ denotes the index of the unit cell. By carefully adjusting the tunable loads of all the unit cells, the desired phase shifts (and amplitude gains) can be achieved thus the reradiated \ac{em} waves can be beamformed toward specified directions. 
As shown in Fig.~\ref{fig_RIS}, different \ac{ris} unit cells, e.g., $Z_{\text{RIS},m}$ and $Z_{\text{RIS},n}$, can be further connected through a reconfigurable impedance  $Z_{\text{RIS},m,n}$, which is referred to as the beyond diagonal \ac{ris} or group/fully-connected \ac{ris} as recently proposed in~\cite{Li2023Beyond,Li2023Beyond2}. 
For clarification, this work uses the classical \ac{ris} with independent unit cells, which is also called the single-connected \ac{ris}.

Each unit cell of the \ac{ris} can be treated as a 1-port network. The voltage waves impinging on the port (denoted as $V_\mathrm{in}$) and reflected from the port (denoted as $V_\mathrm{re}$) are related through the reflection coefficient~$\Theta$ as
\begin{equation}
	V_\mathrm{re} = \Theta V_\mathrm{in}.
\end{equation}
That is, a phase shift $\angle\Theta$ and an amplitude gain $|\Theta|$ are imposed on the incident \ac{em} wave. The reflection coefficient~$\Theta$ is determined by the load impedance~$Z_\mathrm{RIS}$ of the \ac{ris} unit cell through the relationship~\cite{Shen2022Modeling,Pozar2011Microwave}
\begin{equation}\label{eq_ThetaSin}
	\Theta = \frac{Z_\mathrm{RIS}-Z_0}{Z_\mathrm{RIS}+Z_0},
\end{equation}
where $Z_0$ is the reference impedance (or the characteristic impedance) and usually $Z_0=\unit[50]{\Omega}$.
Depending on the real part of the load impedance $Z_\mathrm{RIS}$, various types of RIS can be distinguished as follows.
\begin{itemize}
	\item $\mathrm{Re}(Z_\mathrm{RIS})>0$: This yields an amplitude gain $|\Theta|<1$ and is the case of the most real hardware of the quasi-passive RIS~\cite{Wang2023Wideband}.
	\item $\mathrm{Re}(Z_\mathrm{RIS})=0$: This yields a unit-modulus constraint $|\Theta|=1$ which is widely adopted in the existing theoretical work for simplicity. The unit-modulus reflection coefficient can be realized through a purely reactive RIS load~\cite{Huang2019Reconfigurable,Pan2022An,Shen2022Modeling,Chen2023Multi,Zheng2023Misspecified}.
	\item $\mathrm{Re}(Z_\mathrm{RIS})<0$: This yields an amplitude gain $|\Theta|>1$ which is known as the active \ac{ris}~\cite{Rao2023An}. Several technologies can realize a negative resistance. For example, by imposing proper bias voltage, the tunnel diode can work in the negative differential resistance regions thus leading to an active power gain~\cite{Amato2018Tunneling}.
\end{itemize}


\subsection{A Fabricated Quasi-Passive RIS Prototype}\label{sec_RIS}

Based on the functional principle in Section~\ref{sec_IIA}, a practical wideband 1-bit quantized quasi-passive RIS operating at the 5G mmWave band has been designed and fabricated for the verification of the mutual coupling effect. This RIS prototype utilizes a single-connected network architecture~\cite{Li2023Beyond}. The RIS unit cell design, array synthesis, simulation, and fabricated prototype are demonstrated as follows.

\subsubsection{RIS Unit Cell Design}\label{sec_RISunitcell}

\begin{figure}[t]
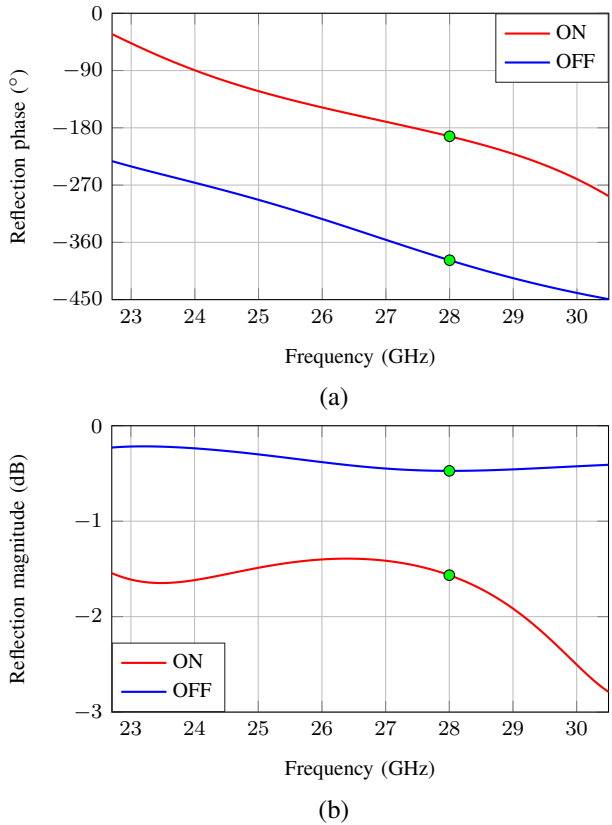

\centering
  \begin{minipage}[b]{1\linewidth}
      \include{figures/RISPhase.tex}
      \vspace{-2.5em}
  \end{minipage}
  (a)\\
  \begin{minipage}[b]{1\linewidth}
      \include{figures/RISMagnitude.tex}
      \vspace{-2.5em}
  \end{minipage}
  (b)
  \caption{
	The designed RIS unit cell performance between ON and OFF states. (a) Phase shift. (b) Amplitude gain.
  }
  \label{fig_RISPhaMag}
\end{figure}

The detailed design methodology for wideband RIS unit element has been investigated in our previous work~\cite{Wang2023Wideband}. The reflection phase and reflection magnitude of the designed unit cell (with \unit[0]{$^\circ$} incidence) when the PIN switch at ON/OFF states are shown in Fig.~\ref{fig_RISPhaMag}.  {Such a reflection performance is realized by employing patches with circular cutouts~\cite{Klionovski2017Physically,Klionovski2019A}, two long vias, one short via, and PIN diodes to excite the unit cell with two adjacent high-order harmonic resonances. Basically, the 1-bit RIS unit cell design has two states when the PIN diodes are switched ON and OFF. Ideally, these two states should have full reflection (magnitude equal to 1 for quasi-passive RIS) and~$\unit[180]{^\circ}$ phase difference for 1-bit phase quantization. However, considering the imperfections of the practical realization, a phase difference of $\unit[180]{^\circ}\pm\unit[20]{^\circ}$ is acceptable, which maintains a decent RIS performance.} We can observe a reflection amplitude larger than $\unit[-2.8]{dB}$ and a phase difference within $\unit[180]{^\circ}\pm\unit[20]{^\circ}$ for the bandwidth from $\unit[22.7]{}$ to $\unit[30.5]{GHz}$. For example, at the frequency of $\unit[28]{GHz}$ (as highlighted in Fig.~\ref{fig_RISPhaMag}), each RIS unit cell has the phase and amplitude responses as follows. 
\begin{equation}
	\angle\Theta =
\begin{cases}
\unit[-193.3]{^\circ} & \text{if } \text{PIN state} = \text{ON},\\
\unit[-388.1]{^\circ} & \text{if } \text{PIN state} = \text{OFF},
\end{cases}
\end{equation}
\begin{equation}
	|\Theta| =
\begin{cases}
\unit[-1.57]{dB} & \text{if } \text{PIN state} = \text{ON},\\
\unit[-0.47]{dB} & \text{if } \text{PIN state} = \text{OFF}.
\end{cases}
\end{equation}
 {Here, the reduced reflection magnitude in the ON state compared to the OFF state is attributed to the PIN diode losses, whereas there are no such losses in the OFF state.}

The designed RIS has a small unit cell size of $0.35\lambda\times 0.35\lambda$ at the center operation frequency of 27.5 GHz. Based on the existing results in the literature, the mutual coupling effect among RIS unit cells becomes severe when the inter-element distance is shorter than half-wavelength~\cite{Qian2021Mutual}. Therefore, the designed RIS prototype should have an observable strong mutual coupling between adjacent RIS cells, suggesting its suitability for the experimental validation of mutual coupling.

\subsubsection{RIS Array Simulation}

\begin{figure}[t]
  \centering
  \includegraphics[width=0.99\linewidth]{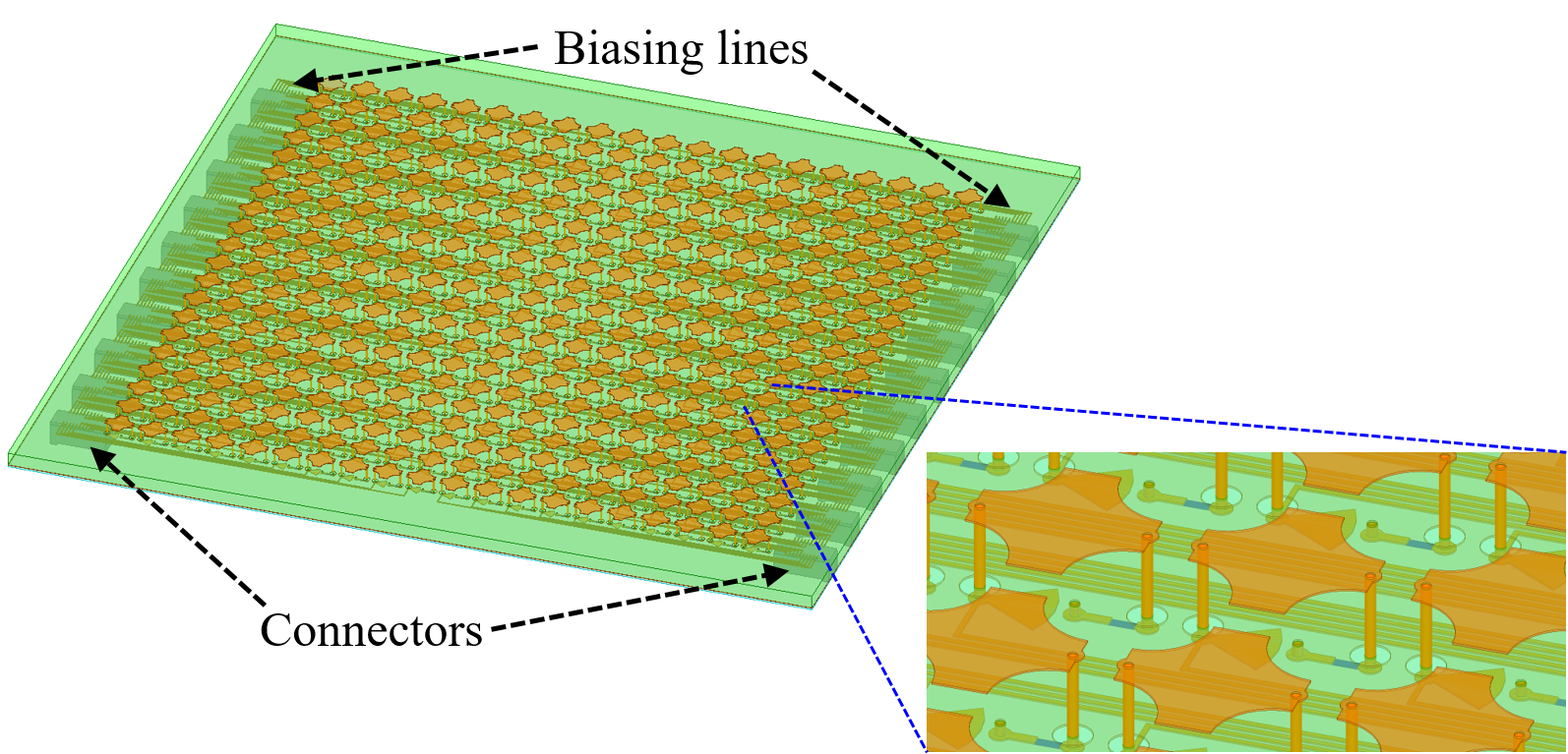}
  \caption{ 
      The practical RIS design. 
    }
  \label{fig_RISconfi}
\end{figure}

\begin{figure}
  \centering
  \begin{minipage}[b]{0.49\linewidth}
  \includegraphics[width=1\linewidth]{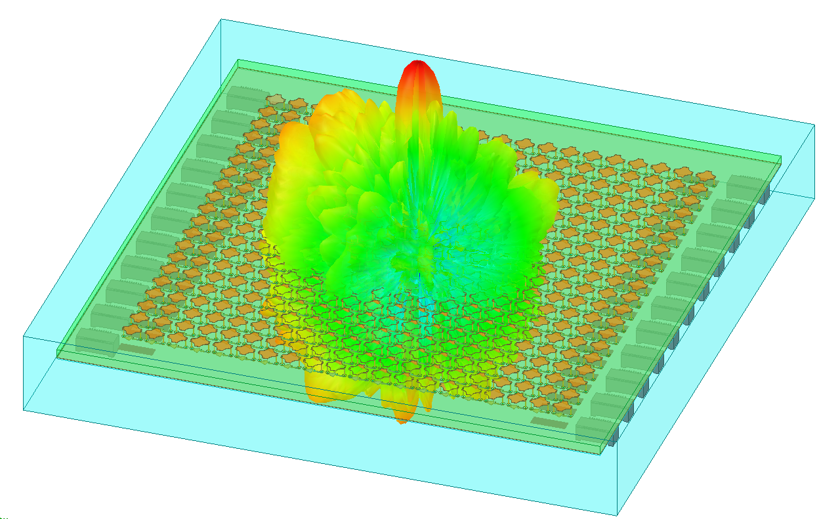}
  \vspace{1em}
  \end{minipage}
  \begin{minipage}[b]{0.49\linewidth}
  \includegraphics[width=1\linewidth]{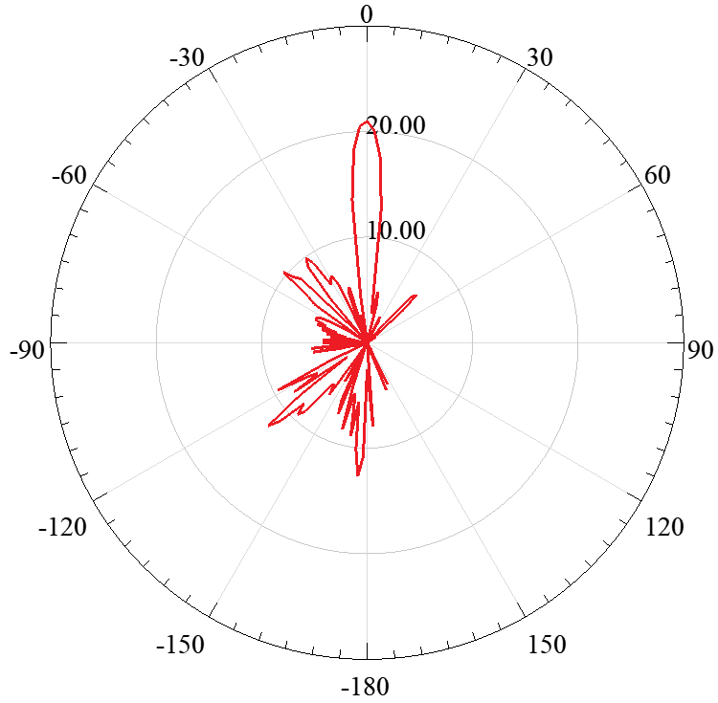}
  \end{minipage}
  \caption{The simulated radiation patterns of the designed RIS with array configuration for $\unit[30]{^\circ}$ illumination and a desired $\unit[0]{^\circ}$--reflection beamforming at $\unit[27.5]{GHz}$.}
  \label{fig_RISsim}
\end{figure}

Before the fabrication and measurement, the radiation performance of the designed RIS can be evaluated through the full-wave simulation based on the finite element method (FEM), 
which captures precise \ac{em} characteristics including mutual coupling. 
For this design, the RIS is composed of 400 unit elements in a 20 × 20 configuration, as demonstrated in Fig.~\ref{fig_RISconfi}. The simulation of the RIS with 400 elements and an incident horn antenna has almost reached the computation limit of the workstation with \unit[256]{GB} RAM. Fig.~\ref{fig_RISsim} illustrates a sample of the simulated 3D radiation pattern along with the RIS and in a polar plot. 
It should be noted that different radiation patterns can be achieved by varying the signal incident angle, RIS beamforming, and the distance between RIS and the incident horn antenna.
However, the mutual coupling parameters, as will be shown in Section~\ref{sec_MCChanModel}, are independent of these system setups and only depend on the physical layout of RIS reflecting elements.
Therefore, the simulated radiation patterns under different system setups can be utilized to train and test the model parameters of the mutual coupling, since it is unable to measure them directly.

\subsubsection{Fabricated Prototype}\label{sec_prototype}

\begin{figure*}[t]
  \centering
  \includegraphics[width=0.99\linewidth]{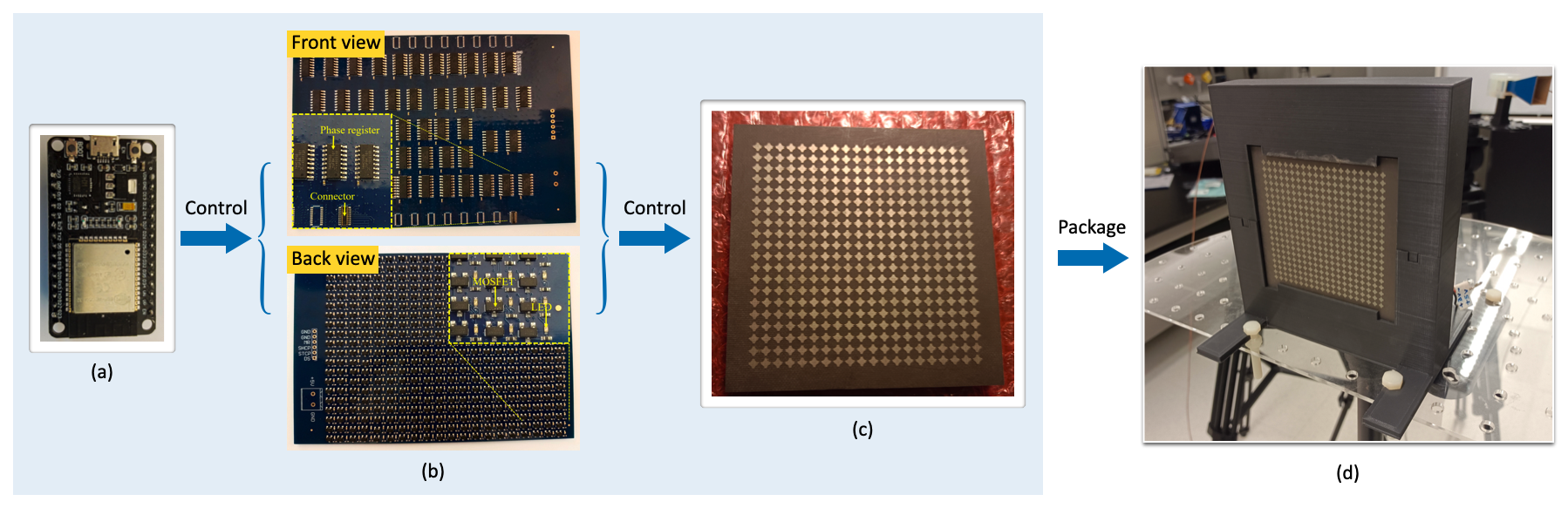}
  \caption{ 
      The fabricated RIS prototype. (a) WROOM-32 module. (b) The designed circuits for the RIS patterns control. (c) The array of the reflective unit cells. (d) The fabricated RIS prototype with integrated control circuits in a package. 
    }
  \label{fig_prototype}
\end{figure*}

Typically, a completely functional RIS design is comprised of a periodic structure with magnitude/phase modulation and a digital circuit for the array pattern generation and control. In this work, the designed control circuits and the packaged RIS prototype are shown in Fig.~\ref{fig_prototype}. 
Specifically, the designed circuit board links with a microcontroller unit (MCU) WROOM-32 module that can compute the array patterns, as illustrated in Fig.~\ref{fig_prototype}-(a).
The circuit is designed with 8-bit phase registers (74HC595D) and PMOS to bias the PIN switches, as shown in Fig.~\ref{fig_prototype}-(b). When the output of the phase register is at the high voltage level, the PIN switch is turned on, thus the RIS unit cell, as demonstrated in Fig.~\ref{fig_prototype}-(c), is in the ON state. Otherwise, the corresponding unit cell is in the OFF state.  
The phase distribution map of the RIS can be visualized through an LED array in the back of the control circuit.  
Fig.~\ref{fig_prototype}-(d) demonstrates the overall fabricated RIS prototype, including the RIS array, control circuit board, connectors, and flexible printed circuit (FPC), which are packaged together through 3D-printed mechanical supports using a Raise3D Pro2 printer with polylactic acid filaments. Note that the function of FPC here is to connect the RIS array with the whole designed circuits inside the package.

\section{RIS-Aided Communication Models}

As a foundation of the subsequent analyses, this section reviews the RIS-aided communication models, including both the conventional model that does not account for RIS mutual coupling~\cite{Liu2021Reconfigurable} and the recently proposed mutual coupling-aware communication model~\cite{Li2023Beyond2}. 
Consider a RIS-aided \ac{mimo} communication with an $M_\mathrm{T}$-antenna transmitter, an $M_\mathrm{R}$-antenna receiver, and a RIS with $N$ unit cells. Let $\vv_\mathrm{R}\in\mathbb{C}^{M_\mathrm{R}\times 1}$ and $\vv_\mathrm{T}\in\mathbb{C}^{M_\mathrm{T}\times 1}$ denote the total voltages at the receiver and the transmitter, respectively.
The wireless channel is represented by the channel matrix~$\Hm\in\mathbb{C}^{M_\mathrm{R}\times M_\mathrm{T}}$,  which builds the connection between $\vv_\mathrm{R}$ and $\vv_\mathrm{T}$ as
\begin{equation}
	\vv_\mathrm{R} = \Hm \vv_\mathrm{T}.
\end{equation}

\subsection{Conventional Channel Model without Mutual Coupling}

A commonly adopted RIS-aided communication channel $\Hm$ can be expressed as~\cite{Liu2021Reconfigurable,ElMossallamy2020Reconfigurable,ChenTutorial2022,Pan2022An,Chen2023Multi,Huang2019Reconfigurable}
\begin{equation}\label{eq_Hconven}
	\Hm = \Hm_\mathrm{RT} + \Hm_\mathrm{RI}\Thetam\Hm_\mathrm{IT},
\end{equation}
where $\Hm_\mathrm{RT}\in\mathbb{C}^{M_\mathrm{R}\times M_\mathrm{T}}$ denotes the Tx-Rx channel, $\Hm_\mathrm{IT}\in\mathbb{C}^{N\times M_\mathrm{T}}$ denotes the Tx-RIS channel, $\Hm_\mathrm{RI}\in\mathbb{C}^{M_\mathrm{R}\times N}$ denotes the RIS-Rx channel, and $\Thetam\in\mathbb{C}^{N\times N}$ is the RIS reflection matrix. Here, $\Thetam$ is a diagonal matrix whose diagonal entries are given by~\eqref{eq_ThetaSin}. The expressions of these channel matrices~$\Hm_\mathrm{RT}$,~$\Hm_\mathrm{RI}$, and~$\Hm_\mathrm{IT}$ will be specified in Section~\ref{sec_ER}.

Although channel model~\eqref{eq_Hconven} is widely utilized, it disregards the interactions between adjacent RIS cells, i.e., the mutual coupling effect. In general, the mutual coupling in an antenna array/RIS can be reasonably ignored if the inter-element distance is large enough (e.g., larger than the half wavelength~\cite{Qian2021Mutual,Zheng2023Impact}). Nonetheless, the significance of mutual coupling becomes increasingly pronounced with the advent of techniques like holographic \ac{mimo} with extremely compact array integration~ {\cite{Pizzo2020Spatially,Gong2024Holographic}}, yielding an inter-element distance much shorter than the half wavelength. Therefore, the mutual coupling-aware communication models are required.

\subsection{Mutual Coupling-Aware Channel Model}\label{sec_MCChanModel}

\begin{figure}[t]
    \centering
    \begin{tikzpicture}
    \node (image) [anchor=south west]{\includegraphics[width=0.9\linewidth]{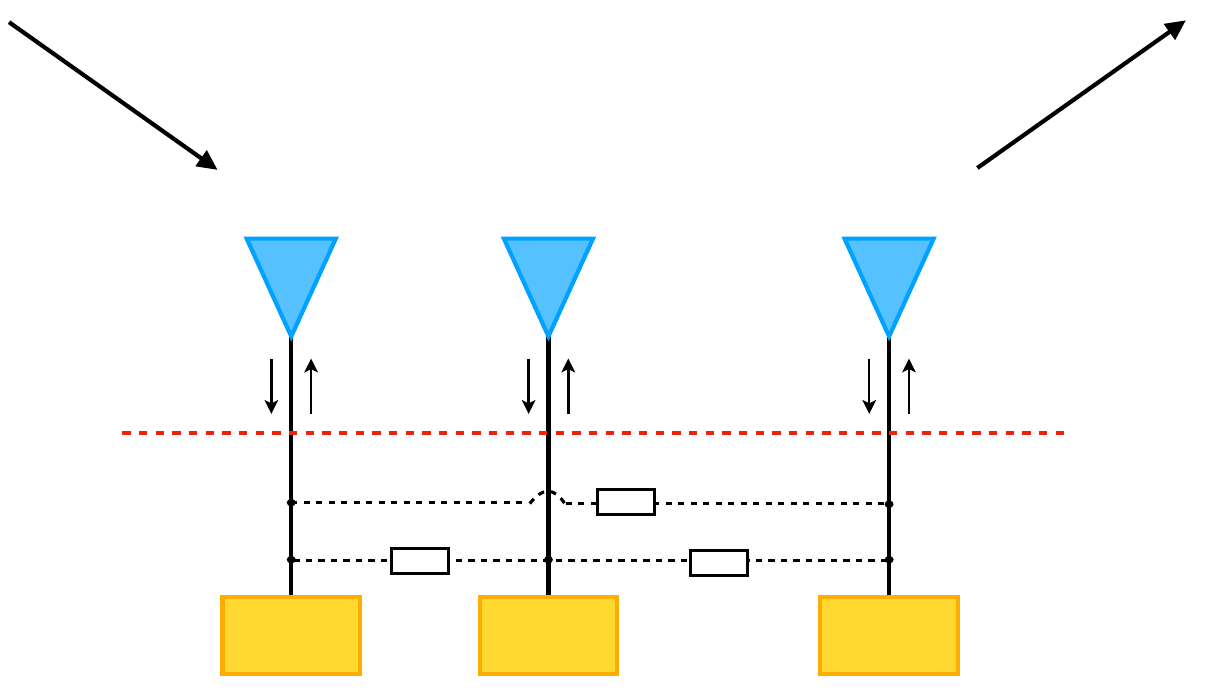}};
    \gettikzxy{(image.north east)}{\ix}{\iy};
	\node at (0.255*\ix,0.158*\iy)[rotate=0,anchor=north]{\scriptsize{$Z_{\text{RIS,1}}$}};
	\node at (0.457*\ix,0.158*\iy)[rotate=0,anchor=north]{\scriptsize{$Z_{\text{RIS,2}}$}};
	\node at (0.735*\ix,0.158*\iy)[rotate=0,anchor=north]{\scriptsize{$Z_{\text{RIS,\textit{N}}}$}};
	\node at (0.36*\ix,0.30*\iy)[rotate=0,anchor=north]{\fontsize{4}{4}\selectfont $Z_{\text{RIS,1,2}}$};
	\node at (0.522*\ix,0.385*\iy)[rotate=0,anchor=north]{\fontsize{4}{4}\selectfont $Z_{\text{RIS,1,\textit{N}}}$};
	\node at (0.6*\ix,0.30*\iy)[rotate=0,anchor=north]{\fontsize{4}{4}\selectfont $Z_{\text{RIS,2,\textit{N}}}$};
	\node at (0.592*\ix,0.13*\iy)[rotate=0,anchor=north]{\dots};
	\draw[red, thick, -stealth, dashed] (0.35*\ix,0.7*\iy) .. controls (0.1*\ix,0.9*\iy) and (0.9*\ix,0.9*\iy) .. (0.65*\ix,0.7*\iy);
     \node at (0.5*\ix,0.75*\iy){\small\red{$\mathbf{S}$}};
     \node at (0.205*\ix,0.5*\iy)[rotate=0,anchor=north]{\scriptsize{$a_{1}$}};
     \node at (0.3*\ix,0.5*\iy)[rotate=0,anchor=north]{\scriptsize{$b_{1}$}};
     \node at (0.41*\ix,0.5*\iy)[rotate=0,anchor=north]{\scriptsize{$a_{2}$}};
     \node at (0.505*\ix,0.5*\iy)[rotate=0,anchor=north]{\scriptsize{$b_{2}$}};
     \node at (0.685*\ix,0.5*\iy)[rotate=0,anchor=north]{\scriptsize{$a_{N}$}};
     \node at (0.78*\ix,0.5*\iy)[rotate=0,anchor=north]{\scriptsize{$b_{N}$}};
     \node at (0.15*\ix,0.9*\iy){\small{$\mathbf{H}_{\mathrm{IT}}$}};
     \node at (0.85*\ix,0.9*\iy){\small{$\mathbf{H}_{\mathrm{RI}}$}};
    \end{tikzpicture}
    \caption{Network representation of a typical RIS-aided wireless communication system, where the RIS can be treated as an $N$-port network. The mutual coupling effect is characterized by the scattering matrix $\Sm$ among RIS reflecting elements.}
    \label{fig_systemNet}
\end{figure}
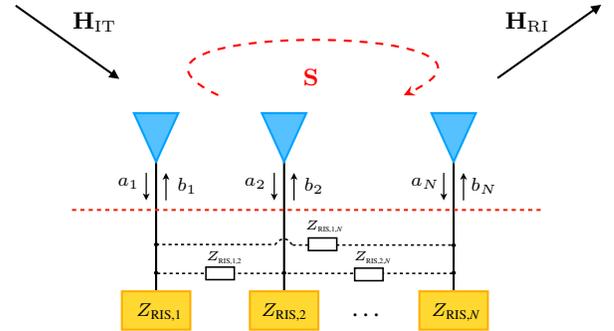

To account for the mutual coupling effect in RIS-aided communication, several novel communication models have been recently proposed. These models can be categorized into two types: the impedance matrix (Z-parameters)-based model~\cite{Gradoni2021End} and the scattering matrix (S-parameters)-based model~\cite{Shen2022Modeling}. It should be noted that based on microwave network theory, these two types of models are essentially equivalent and can be seamlessly converted to each other. Such a relationship has been thoroughly analyzed in~\cite{Li2023Beyond2} and~\cite{Abrardo2023Design}.

In this work, we adopt the scattering matrix-based communication model, which was first reported in~\cite[Eq.~(5)]{Li2023Beyond2}\footnote{ {This work chooses the S-parameter representation instead of the Z-parameter aiming to attain a better compatibility with the conventional wireless channel model.} Note that the transmission scattering matrices $\Sm_\mathrm{RT}$, $\Sm_\mathrm{RI}$, and $\Sm_\mathrm{IT}$ in~\cite[Eq.~(5)]{Li2023Beyond2} are equivalently the channel matrices from the transmitter to receiver, from the transmitter to RIS, and from the RIS to receiver, respectively. This relationship has been proved by~\cite[Eq.~(35)--(39)]{Shen2022Modeling}.} as 
\begin{equation}\label{eq_simpH2}
	\Hm = \Hm_\mathrm{RT}+\Hm_\mathrm{RI}\big(\Thetam^{-1}-\Sm\big)^{-1}\Hm_\mathrm{IT},
\end{equation}
where $\Sm\in\mathbb{C}^{N\times N}$ denotes the scattering matrix of the RIS reflecting element network and thus characterizes the mutual coupling. 
Specifically, the entry $S_{ij}$ in $\Sm$ is the S-parameter between the~$i$-th element and the $j$-th element on the RIS, which indicates the voltage wave measured at the $i$-th element when a unit voltage wave is driven at the $j$-th element~\cite{Pozar2011Microwave}. 
By reciprocity, we have $S_{ij}=S_{ji}$ thus $\Sm$ is a symmetric matrix.
Additionally, the diagonal entry $S_{ii}$ denotes the self-scattering of the $i$-th reflecting element.
This model~\eqref{eq_simpH2} can also be found in, e.g.,~\cite[Eq.~(8)]{Abrardo2023Design} or~\cite[Eq.~(8)]{Di2023Electromagnetic}. 


\section{Scattering Model Training}\label{sec_SMT}
Now, we have two communication models~\eqref{eq_Hconven} and~\eqref{eq_simpH2}. 
However, as can be observed in~\eqref{eq_simpH2}, to utilize this mutual coupling-aware model, we still need to determine the scattering matrix~$\Sm$. 
Since our RIS prototype only reflects \ac{em} waves passively, this scattering matrix $\Sm$ cannot be measured or simulated directly. 
Hence, in this paper, we first propose a model training method that can estimate matrix $\Sm$ based on a single 3D full-wave simulation of the RIS radiation pattern.
Then, based on the estimated scattering matrix $\hat{\Sm}$, we can evaluate and compare the accuracy of the two communication models.
In this section, we first introduce the theoretical calculation of the RIS radiation pattern based on the two communication models. Next, we propose a model training method leveraging such theoretical models and a 3D full-wave simulation of the RIS radiation pattern. 


\subsection{Theoretical Radiation Pattern Calculation}\label{sec_RP_cal}

 Since our measurement facility with horn Tx and Rx antennas form a \ac{siso} setup, we now present the calculation of the RIS radiation pattern based on the \ac{siso} version of~\eqref{eq_Hconven} and~\eqref{eq_simpH2}. Aiming to evaluate the RIS mutual coupling, we focus on the Tx-RIS-Rx channel only and block the Tx-Rx path that is unrelated to RIS. 

The radiation pattern refers to a function of the radiation properties of an array as a function of the directional coordinates~\cite{Balanis2016Antenna}.\footnote{Radiation properties include power flux density, radiation intensity, field strength, directivity, phase or polarization. By default, this paper refers to the radiation pattern as the amplitude field pattern.}
By fixing the position of the Tx horn antenna, we can calculate the radiation pattern of the Tx$+$RIS system.
We denote the directional coordinates by the \ac{aod} $\thetav=[\theta_\mathrm{az},\theta_\mathrm{el}]^\TT$ departing from the RIS, which consists of an azimuth angle~$\theta_\mathrm{az}$ and an elevation angle~$\theta_\mathrm{el}$. In this paper, the elevation angle is defined as the angle between the departure direction and the normal direction of the array plane, as utilized in~\cite{Balanis2016Antenna}.

\subsubsection{Radiation Pattern Calculation Based on~\eqref{eq_Hconven}}

Based on the conventional communication model~\eqref{eq_Hconven}, different RIS unit cells reflect signals independently without interaction. Hence, the (amplitude) radiation pattern in linear scale can be calculated by~\cite[Eq. (4.7)]{Nayeri2018Reflectarray}
\begin{align}\label{eq_E}
	&E(\thetav)\\ 
	=&\Big|\!\sum_{n=1}^N \Theta_{n,n}\frac{\cos^{q_\mathrm{e}}(\theta_\mathrm{el})\cos^{q_\mathrm{f}}(\theta_{\mathrm{T},n})}{\|\pv_n-\pv_\mathrm{T}\|_2}e^{jk_0(\pv_n^\TT\uv(\thetav)-\|\pv_n-\pv_\mathrm{T}\|_2)}\Big|,\notag
\end{align}
where $k_0={2\pi}/{\lambda}$ is the wavenumber, $\pv_n\in\mathbb{R}^3$ denotes the position of the $n$-th RIS element,~$\pv_\mathrm{T}\in\mathbb{R}^3$ denotes the position of the Tx horn,  and $\uv(\thetav)$ denotes the unit directional vector of the \ac{aod}~$\thetav$. All these positions and vectors are expressed in the body coordinate system of the RIS. Here, $\lambda=c/f$ denotes the \ac{em} wavelength, where $f$ is the signal frequency and $c$ is the speed of light. Besides,~$\theta_{\mathrm{T},n}$ denotes the elevation angle of the \ac{aod} from Tx to the~$n$-th RIS unit cell expressed in the coordinate system of the Tx horn, and $q_\mathrm{e}$ and $q_\mathrm{f}$ are the feed power patterns which determine the radiation directivities of the RIS unit cell and Tx horn antenna, respectively. 
In addition, $\Theta_{n,n}$ denotes the reflection coefficient of the $n$-th RIS unit cell, i.e., the $n$-th diagonal entry of $\Thetam$.

We can reform~\eqref{eq_E} into a more compact form as
\begin{equation}\label{eq_Econv}
	E(\thetav) = \big|\hv_{\mathrm{out}}^\TT(\thetav)\Thetam\hv_\mathrm{in}\big|,
\end{equation}
where the entries in $\hv_\mathrm{in}\in\mathbb{C}^{N\times 1}$ and $\hv_{\mathrm{out}}(\thetav)\in\mathbb{C}^{N\times 1}$ are given by 
\begin{align}
	\label{eq_hIT} h_{\mathrm{in},n} &= \frac{\cos^{q_\mathrm{f}}(\theta_{\mathrm{T},n})}{\|\pv_n-\pv_\mathrm{T}\|_2}e^{-jk_0\|\pv_n-\pv_\mathrm{T}\|_2},\\
	\label{eq_hAoD} h_{\mathrm{out},n}(\thetav) &= \cos^{q_\mathrm{e}}(\theta_\mathrm{el})e^{jk_0\pv_n^\TT\uv(\thetav)}.
\end{align}

\subsubsection{Radiation Pattern Calculation Based on~\eqref{eq_simpH2}}

By replacing the RIS response $\Thetam$ with $(\Thetam^{-1}-\Sm)^{-1}$ according to~\eqref{eq_simpH2}, the mutual coupling-aware RIS radiation pattern can be computed as
\begin{equation}\label{eq_rMC}
	{E}(\thetav) = \big|\hv_{\mathrm{out}}^\TT(\thetav)\big(\Thetam^{-1}-\Sm\big)^{-1}\hv_\mathrm{in}\big|.
\end{equation} 

For clarification, we denote the radiation pattern calculated based on~\eqref{eq_rMC} as~$E(\thetav;\Sm)$, showing it is a function of scattering matrix $\Sm$.
In contrast, we denote the radiation pattern calculated through~\eqref{eq_Econv} as~$E(\thetav;\mathbf{0})$ for distinction.
Besides, radiation pattern of an array is usually normalized with respect to their maximum value, yielding normalized radiation patterns. To do so, we define the normalized version of~\eqref{eq_Econv} and~\eqref{eq_rMC} as
\begin{align}
	E_\mathrm{n}(\thetav;\mathbf{0}) &=  E(\thetav;\mathbf{0})/\big(\max_{\thetav} E(\thetav;\mathbf{0})\big),\\
	E_\mathrm{n}(\thetav;\Sm) &=  E(\thetav;\Sm)/\big(\max_{\thetav} E(\thetav;\Sm)\big).\label{eq_Entilde}
\end{align}

\subsection{Scattering Parameters Training}\label{sec_SPT}

\begin{figure*}[t]
  \centering
  \includegraphics[width=0.99\linewidth]{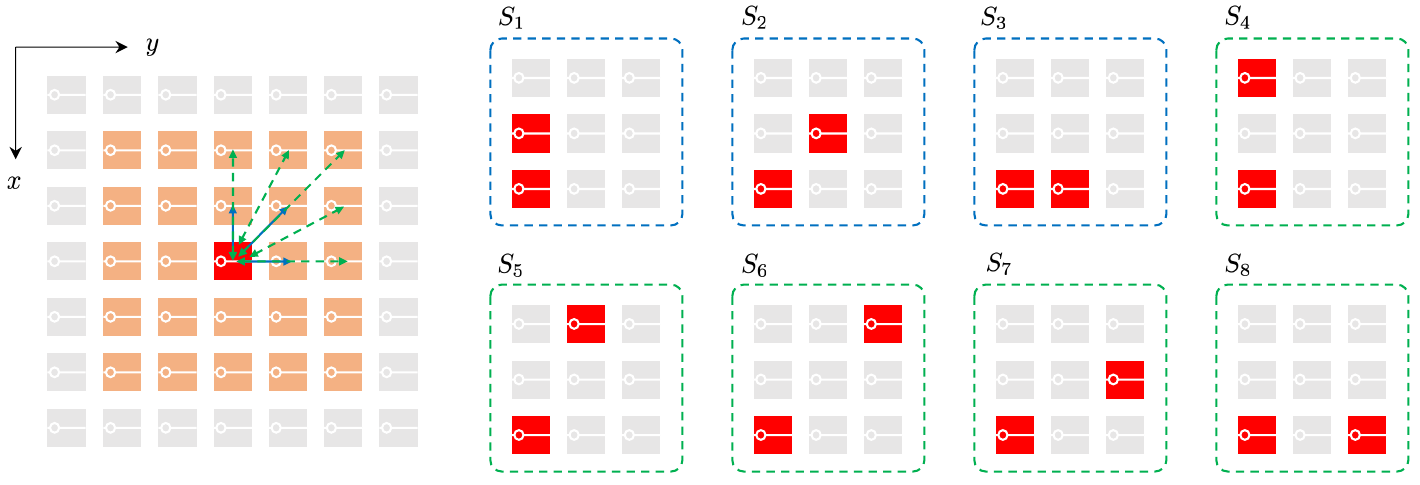}
  \caption{ 
      The illustration of different types of S-parameter based on geometric configurations. For each RIS reflecting element, we only consider the coupling with its nearest 24 neighboring elements. As depicted, our RIS unit cells are folded symmetric along the $y$-axis, but not rotationally symmetric. Therefore, based on geometric relationships, there are only 8 different values of S-parameters in $\Sm$, which we represent as $S_1,S_2,\dots,S_8$, corresponding to the cases shown in the figure. In addition, we use $S_0$ to denote the self-scattering parameter.
    }
  \label{fig_S028}
\end{figure*}

Based on the derivation in Section~\ref{sec_RP_cal}, we now design a training procedure to estimate the unknown scattering matrix~$\Sm$ based on~\eqref{eq_rMC}. 
As mentioned in Section~\ref{sec_MCChanModel}, matrix~$\Sm$ is symmetric, indicating that the \ac{dof} of~$\Sm$ is ~$N(N+1)/2$.
Considering the geometry of the \ac{upa} of RIS elements, the \ac{dof} of $\Sm$ can be further reduced.

Typically, introducing a third element between two existing element will result in an \ac{em} interaction characterized by indirect coupling, which is generally exceedingly weak. Moreover, this attenuation intensifies with an increasing number of elements in between, implying that the mutual coupling between two elements separated by multiple intervening elements is negligible. Therefore, in this study, we only consider the mutual coupling between element pairs with 0 or 1 intervening element, while ignoring the rest, as shown in Fig.~\ref{fig_S028}. Furthermore, we note that the unit cells in our RIS prototype exhibit fold symmetry along the $y$-axis rather than rotational symmetry. In consideration of such geometric configuration, we can summarize that there are only 8 different values of S-parameters in $\Sm$, which we denote as $S_1,S_2,\dots,S_8$, as depicted in Fig.~\ref{fig_S028}. Additionally, we denote the self-scattering parameter as $S_0$. Hence, we can approximately express the scattering matrix $\Sm$ as 
\begin{equation}
	\Sm \approx \sum_{i=0}^8S_i\Am_i,
\end{equation}
 {where $\Am_i$ is the known support matrix of the value~$S_i$ in~$\Sm$ according to array configuration. Namely, the entry at the $j$-th row and $k$-th column is $A_{i,jk}=1$ for all $S_{jk}=S_i$; otherwise, $A_{i,jk}=0$. For example, the support matrix for the self-coupling $S_0$ is $\Am_0=\mathbf{I}_{N^2}$.}
This expression suggests the \ac{dof} of $\Sm$ is reduced to 9, which is independent of the RIS size $N$.
In other words, the scattering matrix $\Sm$ can be determined by these~9 parameters~$\{S_0,\dots,S_8\}$.  {In addition, given that the adopted RIS prototype exhibits robust and stable unit cell magnitude and phase performance within $0-50^\circ$ incidences at \unit[27.5]{GHz}~\cite{Wang2023Wideband}, we assume for simplicity in our analysis that these 9 parameters are independent of the incident angle of \ac{em} waves.}

According to~\eqref{eq_Entilde}, the mutual coupling-aware normalized radiation pattern~$E_\mathrm{n}(\thetav;\Sm)$ is a function of $\Sm$ and thus is a function of parameters $\{S_0,\dots,S_8\}$. Suppose there is a normalized radiation pattern obtained through a 3D full-wave simulation based on the actual RIS prototype, denoted as $\bar{E}_\mathrm{n}(\thetav)$. We sample this simulated pattern over \ac{aod} $\thetav$, thus acquiring the discrete version of the simulated pattern as $\bar{E}_\mathrm{n}(\thetav_\ell),\ell=1,\dots,L_\St$. Here,~$L_\St$ denotes the number of samples.  {Based on the RIS radiation prediction formula~\eqref{eq_Entilde} and the obtained samples of the actual radiation pattern, we can design an optimization problem to estimate the scattering parameters $\{S_0,\dots,S_8\}$. The objective is to optimize these scattering parameters to minimize the discrepancy between the predicted radiation pattern and the actual radiation pattern. Consequently, a least-squares problem is formulated as follows:
\begin{align}
	&\{\hat{S}_0,\!\dots,\!\hat{S}_8\} \notag \\
	&=\!\arg\min_{S_0,\dots,S_8}\sum_{\ell=1}^{L_\mathrm{S}}\big(\frac{E(\bm{\theta}_\ell;\!S_0,\!\dots,\!S_8\!)}{\max\limits_\ell E(\bm{\theta}_\ell;\!S_0,\!\dots,\!S_8\!)}-\bar{E}_\mathrm{n}(\bm{\theta}_\ell)\big)^2. \label{eq_min}
\end{align}}
Subsequently, the scattering matrix can be recovered as 
\begin{equation}
	\hat{\Sm} = \sum_{i=0}^8\hat{S}_i\Am_i.
\end{equation}
 {In the following subsection, we present a solution to~\eqref{eq_min} by reformulating the optimization problem based on the Neumann series expansion.

\subsection{Solving~\eqref{eq_min} via Neumann Series Approximation}\label{sec:NSE}
Due to the matrix inversion in~\eqref{eq_rMC} and the normalization operation in~\eqref{eq_min}, the objective function in~\eqref{eq_min} is non-convex and non-differentiable. Although gradient-based iterative algorithms can be applied using the numerical gradient (e.g., through finite difference approximation), the local-minimum issue is unavoidable, and there is no guarantee that such algorithms can converge to the global minimum. To this end, here we propose a more tractable method to estimate the scattering parameters by reformulating~\eqref{eq_min} based on the Neumann series expansion.

The Neumann series expansion~\cite{Ortega2013Matrix} reads
\begin{equation}\label{eq:Neumann}
    (\Thetam^{-1}-\Sm)^{-1}\!=\!\Big(\sum_{n=0}^\infty \big(\Thetam\Sm\big)^n\Big)\Thetam
   \!=\! \Thetam\! +\! \Thetam\Sm\Thetam\! +\! \Thetam\Sm\Thetam\Sm\Thetam \!+\! \cdots,
\end{equation}
if all the eigenvalues of~$\Thetam\Sm$ are within the unit circle, i.e.,~$|\lambda_i(\Thetam\Sm)|<1$, $\forall\ i=1,\dots,N$.\footnote{We will show in Section~\ref{sec:MT} that this condition can be satisfied in our RIS prototype.} Therefore, based on~\eqref{eq_rMC} and~\eqref{eq:Neumann}, we have
\begin{align}
	{E}(\thetav) &= \big|\hv_{\mathrm{out}}^\TT(\thetav)\big(\Thetam^{-1}-\Sm\big)^{-1}\hv_\mathrm{in}\big|,\\
	&\overset{(a)}{\approx} \big|\hv_{\mathrm{out}}^\TT(\thetav)\big(\Thetam+\Thetam\Sm\Thetam\big)\hv_\mathrm{in}\big|, \label{eq:appro}
\end{align}
where step~(a) is obtained by only preserving the first two terms of the Neumann series expansion. 

Then, we can reformulate the objective function in~\eqref{eq_min} as 
\begin{multline}
f(S_0,\dots,S_8,\varepsilon) = \Big\|\Big(\Hm_{\mathrm{out}}^\TT\big(\Thetam\!+\!\Thetam\big(\sum_{i=0}^8S_i\Am_i\big)\Thetam\big)\hv_\mathrm{in}\Big)\!\\ \odot\!\Big(\Hm_{\mathrm{out}}^\HH\big(\Thetam^*\!+\!\Thetam^*\big(\sum_{i=0}^8S_i^*\Am_i\big)\Thetam^*\big)\hv_\mathrm{in}^*\Big)\!-\!\varepsilon\bar{\ev}\Big\|_2,\label{eq:f}
\end{multline}
where~$\varepsilon\in\mathbb{R}^+$ is an auxiliary variable counteracting the unknown normalization scale,~$\odot$ denotes Hadamard (elementwise) product,~$\Hm_\mathrm{out}\overset{\Delta}{=}[\hv_\mathrm{out}(\thetav_1),\dots,\hv_\mathrm{out}(\thetav_{L_\St})]$, and $\bar{\ev}\overset{\Delta}{=}[\bar{E}^2_\mathrm{n}(\bm{\theta}_1),\dots,\bar{E}^2_\mathrm{n}(\bm{\theta}_{L_\St})]^\TT$. Subsequently, we can estimate the scattering parameters by solving the following optimization problem:
\begin{equation}\label{eq:opt2}
	\begin{aligned}
	\min_{S_0,\dots,S_8,\varepsilon}\ &f(S_0,\dots,S_8,\varepsilon),\\
	\mathrm{s. t.}\quad &S_0,\dots,S_8\in\mathbb{C},\ \varepsilon\in\mathbb{R}^+.
	\end{aligned}
\end{equation}
This is a convex optimization problem, the convexity of which is presented in Appendix. By applying a gradient descent algorithm, the scattering parameters~$S_0,\dots,S_8$ can be estimated. Note that the direction where the function~$f$ decreases fastest is given by~$-\partial f/\partial [S_0^*,\dots,S_8^*,\varepsilon]^\TT$~\cite[Theorem 3.4]{Hjorungnes2011Complex}. 
}

\section{Simulation Results}
This section first conducts the model training procedure as described in Section~\ref{sec_SPT} and Section~\ref{sec:NSE}, and then assesses the accuracy of the trained model via full-wave simulation.

\subsection{Model training}\label{sec:MT}
Before conducting the estimation process proposed in Section~\ref{sec:NSE}, we first perform a full-wave simulation based on a similar transmitting \ac{mimo} setup to obtain a set of rough estimates of the unknown scattering parameters, which can serve as (i) the initialization of the iterative gradient descent method and (ii) an additional regularization term to the objective function~$f(S_0,\dots,S_8,\varepsilon)$ in~\eqref{eq:f} to mitigate over-fitting.

\subsubsection{Simulation of Transmitting MIMO Antennas}

\begin{figure}[t]
  \centering
  \includegraphics[width=0.99\linewidth]{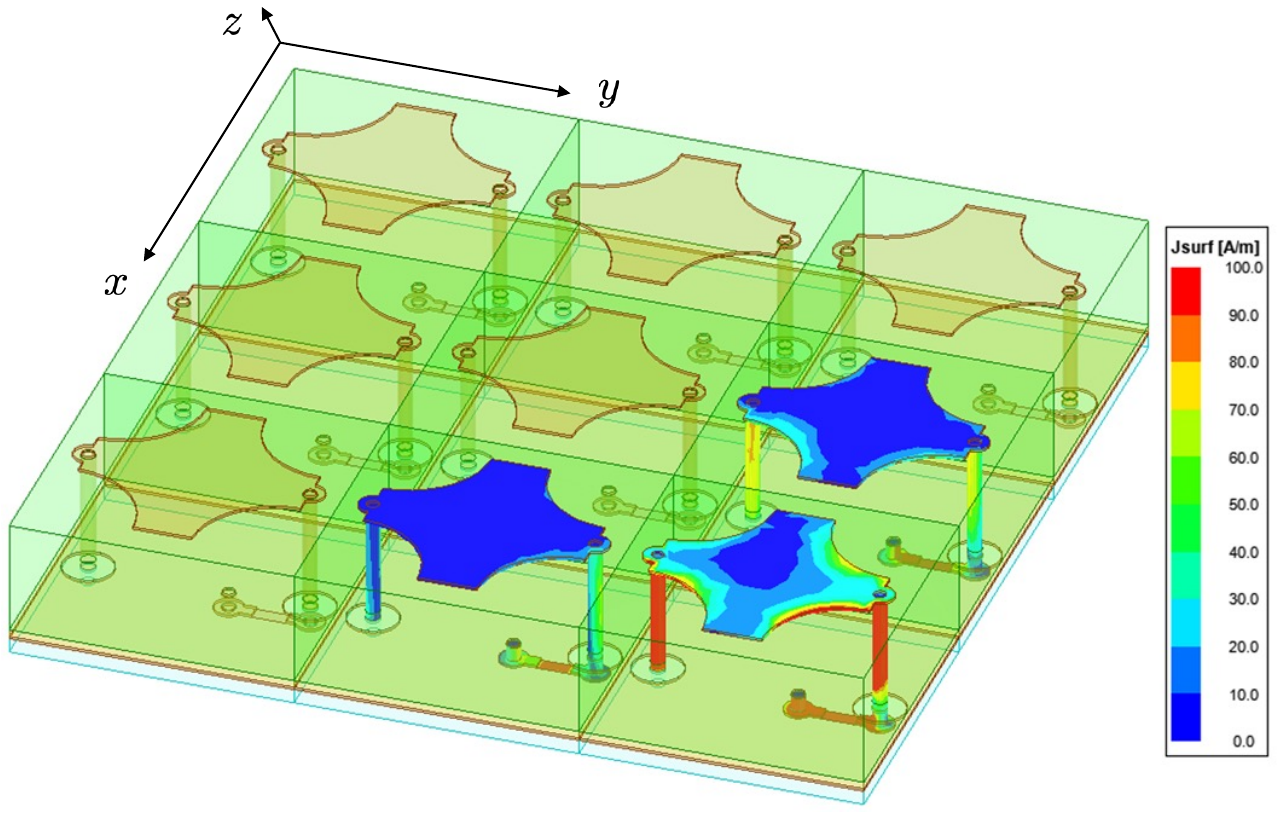}
  \caption{ 
      The simulated antenna model of $3\times 3$ \ac{mimo} configuration, where the surface current distributions on the 3 chosen antenna cells are plotted.
    }
  \label{fig_MIMO}
\end{figure}

\begin{table}[t]
  \renewcommand{\arraystretch}{1.3}
  \begin{center}
  \caption{Simulated Scattering Parameters in MIMO Configuration}\vspace{-0.5em}
  \label{tab1}
  \begin{tabular}{  c !{\vrule width1pt} c !{\vrule} c }
    \Xhline{1pt}
    Scattering Parameter & Magnitude (dB) & Phase (degree) \\
    \Xhline{1pt}
    $\tilde{S}_0$ & $-9.5215$ & $-16.4$ \\
    \Xhline{1pt} 
    $\tilde{S}_1$ & $-15.1365$ & $146.8$ \\
    \hline 
    $\tilde{S}_2$ & $-18.9885$ & $54.4$ \\
    \hline 
    $\tilde{S}_3$ & $-19.1683$ & $40.0$ \\
    \Xhline{1pt}
    $\tilde{S}_4$ & $-25.8747$ & $36.5$ \\
    \hline 
    $\tilde{S}_5$ & $-21.5950$ & $6.5$ \\
    \hline 
    $\tilde{S}_6$ & $-28.4459$ & $-142.4$ \\
    \hline 
    $\tilde{S}_7$ & $-24.9849$ & $-107.4$ \\
    \hline 
    $\tilde{S}_8$ & $-25.6949$ & $-84.5$ \\
    \Xhline{1pt}
    \end{tabular}
\end{center}
\end{table}

As mentioned in Section~\ref{sec_SMT}, the scattering matrix~$\Sm$ for the RIS cannot be measured or simulated directly since RIS is excited through incident EM waves instead of direct feeding ports.  
Nevertheless, from the \ac{em} coupling perspective, the mutual coupling of the reflective RIS elements and the transmitting \ac{mimo} antennas should possess a similar tendency. Hence, a similar antenna model with \ac{mimo} configuration can be simulated using commercial full-wave simulation software such as Ansys HFSS, to obtain an approximated scattering matrix to initialize~\eqref{eq:opt2}. We simulate a $3\times 3$ \ac{mimo} antenna array with a similar physical layout as RIS using Ansys HFSS, as presented in Fig.~\ref{fig_MIMO}.\footnote{A $3\times 3$ antenna array includes all values of $S_0,S_1,\dots,S_8$ shown in Fig.~\ref{fig_S028}.  {Our tests indicate that increasing the array size does not significantly affect the values of these S-parameters.}}

The simulated results of the scattering parameters among such a \ac{mimo} antenna array are listed in Table~\ref{tab1}. 
To be clear, we denote these simulated \ac{mimo} scattering parameters as $\tilde{S}_0,\tilde{S}_1,\dots,\tilde{S}_8$ and the corresponding scattering matrix as $\tilde{\Sm}=\sum_{i=0}^8\tilde{S}_i\Am_i$.
By observing the data in Table~\ref{tab1}, insights are obtained that the adjacent coupling in the $x$-axis direction is stronger than that in the $y$-axis direction, since $|S_1|>|S_3|$. This phenomenon results from the non-rotational symmetry of the antenna structure. 
To further validate and better understand this difference. The surface current distribution of the \ac{mimo} antenna configuration is also demonstrated in Fig.~\ref{fig_MIMO}. Here, the unit in the right-bottom corner is excited, leading to the induction of currents in neighboring units through the coupling effect. Therefore, the EM coupling along the $x$ and $y$-axes can be intuitively observed through the surface current distribution on adjacent unit cells. It is clear that the coupling along the $x$-axis results in a stronger coupled surface current than that along the $y$-axis, coinciding with the simulated results in Table~\ref{tab1}. These observations imply that the mutual coupling in the RIS can also exhibit such a trend.

\subsubsection{RIS Model Training}

\begin{figure*}[t]
    \centering
    \begin{tikzpicture}
    \node (image) [anchor=south west]{\includegraphics[width=1\linewidth]{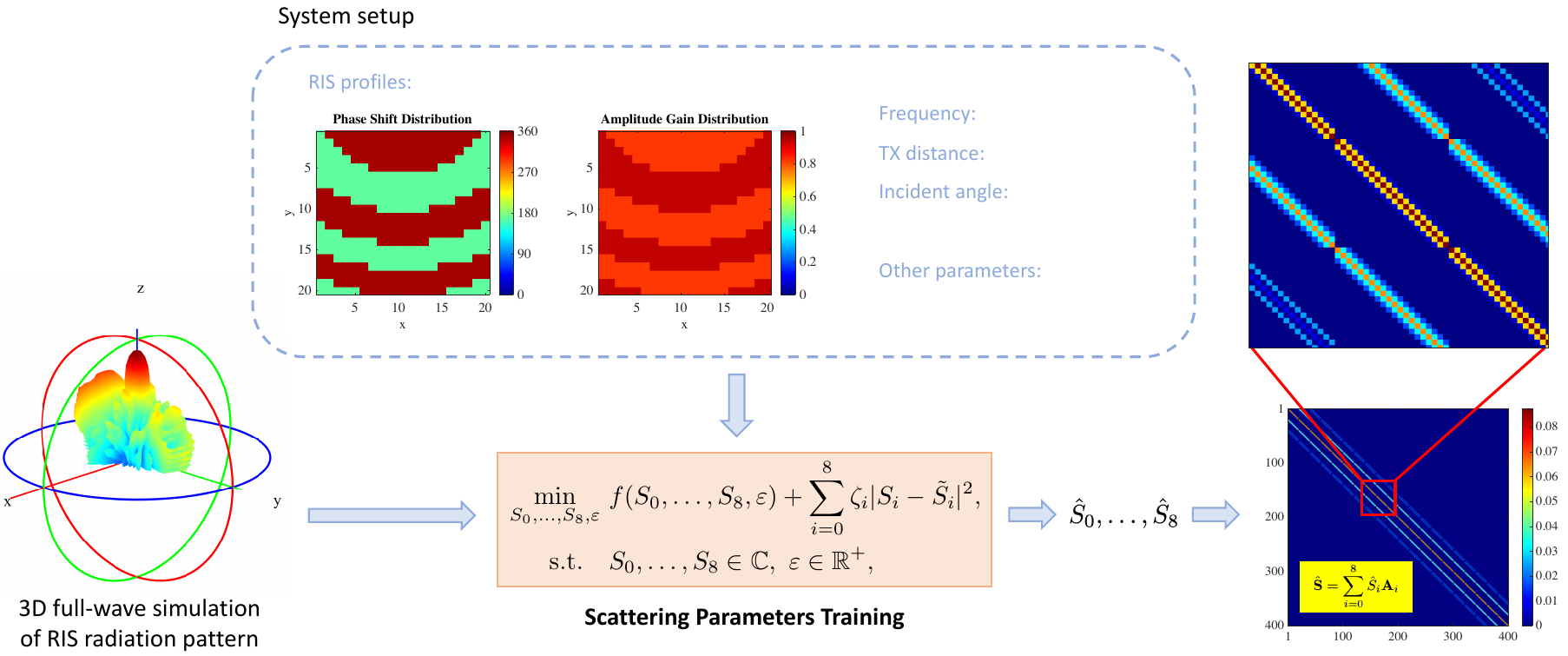}};
    \gettikzxy{(image.north east)}{\ix}{\iy};
     \node at (0.653*\ix,0.815*\iy){\scriptsize{\unit[27.5]{GHz}}};
     \node at (0.653*\ix,0.757*\iy){\scriptsize{\unit[0.18]{m}}};
     \node at (0.67*\ix,0.65*\iy){\scriptsize{$\phi_{\mathrm{az}}=90^\circ,\ \phi_{\mathrm{el}}=20^\circ$}};
     \node at (0.66*\ix,0.53*\iy){\scriptsize{$q_{\mathrm{e}}=0.5,\ q_{\mathrm{f}}=34$}};
     \node at (0.25*\ix,0.29*\iy){\scriptsize{$\{\bar{E}_\mathrm{n}(\thetav_\ell)\}_{\ell=1}^{L_\St}$}};
     \node at (0.25*\ix,0.33*\iy){\scriptsize{Sample}};
    \end{tikzpicture}
    \caption{The scattering parameters training procedure based on a single 3D full-wave simulation of RIS radiation pattern.}
    \label{fig_training}
\end{figure*}
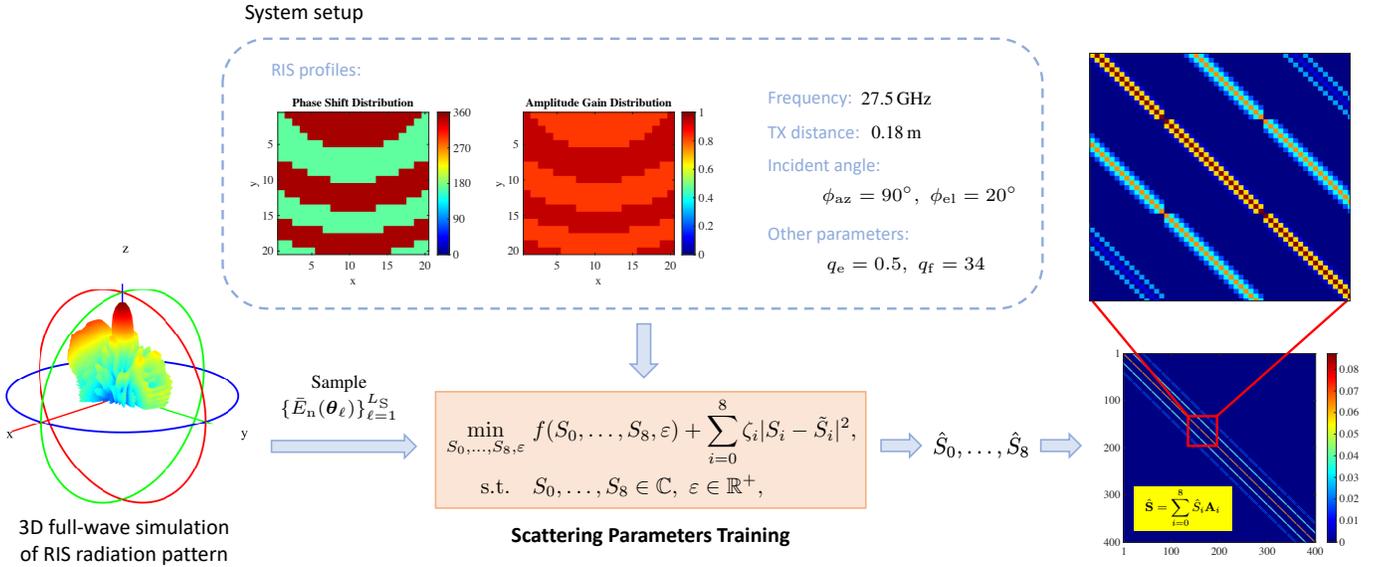

 {
Now we conduct the estimation of the S-parameters based on the method presented in Section~\ref{sec_SPT} and~\ref{sec:NSE}. Using the simulated~$\tilde{\Sm}$ in Table~\ref{tab1}, we can easily verify that the condition $|\lambda_i(\Thetam\tilde{\Sm})|<1$, $\forall\ i=1,\dots,N$ holds. Thus, the Neumann series expansion in~\eqref{eq:Neumann} is valid based on the fact that the true~$\Sm$ is close to~$\tilde{\Sm}$. Besides, Fig.~\ref{Fig_NSAE} plots the truncation error of this Neumann series expansion. It is shown that by considering only the first two terms, we can obtain an approximation with a truncation error of approximately~$0.018\%$ only, which justifies the accuracy of the approximation in~\eqref{eq:appro}.

\begin{figure}[t]
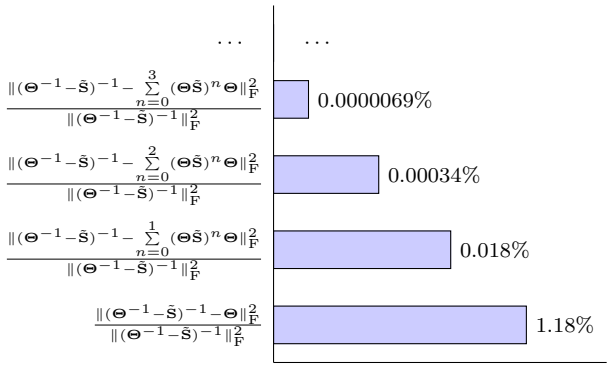

    \centering
    \renewcommand{\bcfontstyle}{\footnotesize}
	\begin{bchart}[plain,max=80,width=0.5\linewidth]
		\bcbar[label=$\cdots\quad$,value=$\quad\cdots$]{0}
		\smallskip
        \bcbar[label={$\frac{\|({\tiny\Thetam}^{-1}-\tilde{\Sm})^{-1}-\sum\limits_{n=0}^3({\tiny\Thetam}\tilde{\Sm})^n{\tiny\Thetam}\|_\Ft^2}{\|({\tiny\Thetam}^{-1}-\tilde{\Sm})^{-1}\|_\Ft^2}$},value=$0.0000069\%$]{8.3924}
            \medskip
        \bcbar[label={$\frac{\|({\tiny\Thetam}^{-1}-\tilde{\Sm})^{-1}-\sum\limits_{n=0}^2({\tiny\Thetam}\tilde{\Sm})^n{\tiny\Thetam}\|_\Ft^2}{\|({\tiny\Thetam}^{-1}-\tilde{\Sm})^{-1}\|_\Ft^2}$},value=$0.00034\%$]{25.2546}
            \medskip
        \bcbar[label={$\frac{\|({\tiny\Thetam}^{-1}-\tilde{\Sm})^{-1}-\sum\limits_{n=0}^1({\tiny\Thetam}\tilde{\Sm})^n{\tiny\Thetam}\|_\Ft^2}{\|({\tiny\Thetam}^{-1}-\tilde{\Sm})^{-1}\|_\Ft^2}$},value=$0.018\%$]{42.5429}
            \medskip
        \bcbar[label={$\frac{\|({\tiny\Thetam}^{-1}-\tilde{\Sm})^{-1}-{\tiny\Thetam}\|_\Ft^2}{\|({\tiny\Thetam}^{-1}-\tilde{\Sm})^{-1}\|_\Ft^2}$},value=$1.18\%$]{60.7155}
    \end{bchart}
  	\caption{The truncation error of the Neumann series expansion~\eqref{eq:Neumann} based on~$\tilde{\Sm}$ in Table~\ref{tab1}. }
\label{Fig_NSAE}
\end{figure}

Besides justifying the accuracy of the Neumann series-based approximation, the simulated~$\tilde{\Sm}$ can further benefit our estimation by providing a regularization penalty. Specifically, we rewrite~\eqref{eq:opt2} as 
\begin{equation}\label{eq:opt3}
	\begin{aligned}
	\min_{S_0,\dots,S_8,\varepsilon}\ &f(S_0,\dots,S_8,\varepsilon) + \sum_{i=0}^8\zeta_i|S_i-\tilde{S}_i|^2,\\
	\mathrm{s. t.}\quad &S_0,\dots,S_8\in\mathbb{C},\ \varepsilon\in\mathbb{R}^+,
	\end{aligned}
\end{equation}
which is still a convex problem and~$\zeta_i$ controls the importance/trust of the each regularization term.
}
Utilizing the simulated data in Table~\ref{tab1} as an initialization, this convex optimization problem can be solved using the gradient descent method. The observations $\bar{E}_\mathrm{n}(\thetav_\ell),\ell=1,2,\dots,L_\St$, are obtained through a single 3D full-wave simulation of RIS radiation pattern using Ansys HFSS. Here, we sample the 3D pattern uniformly such that $L_\St=91\times 91$. This simulation is conducted at frequency of~$\unit[27.5]{GHz}$, where the Tx horn is placed at a distance of $\unit[0.18]{m}$ to the RIS with incident angle $\phi_\mathrm{az}=90^\circ,\ \phi_\mathrm{el}=20^\circ$.\footnote{ {The designed RIS prototype can only operate in a single polarization, where the incident wave polarization must be along the $y$-axis, i.e.,~$\phi_\mathrm{az}=90^\circ$, according to the coordinate system plotted in Fig.~\ref{fig_MIMO}.}} The RIS phase shifts and amplitude gains are assigned following the beamforming method in~\cite{Wang2023Wideband}. 
The overall model training procedure and results are visualized in Fig.~\ref{fig_training}, and the values of the estimated scattering parameters~$\hat{S}_0,\hat{S}_1,\dots,\hat{S}_8$ are listed in Table~\ref{tab2}.

\begin{table}[t]
  \renewcommand{\arraystretch}{1.3}
  \begin{center}
  \caption{Estimated Scattering Parameters in Fig.~\ref{fig_training}}\vspace{-0.5em}
  \label{tab2}
  \begin{tabular}{  c !{\vrule width1pt} c !{\vrule} c }
    \Xhline{1pt}
    Scattering Parameter & Magnitude (dB) & Phase (degree) \\
    \Xhline{1pt}
    $\hat{S}_0$ & $-10.5914$ & $25.5$ \\
    \Xhline{1pt} 
    $\hat{S}_1$ & $-11.9197$ & $121.5$ \\
    \hline 
    $\hat{S}_2$ & $-14.8965$ & $-117.0$ \\
    \hline 
    $\hat{S}_3$ & $-12.4197$ & $-14.3$ \\
    \Xhline{1pt}
    $\hat{S}_4$ & $-20.7811$ & $83.8$ \\
    \hline 
    $\hat{S}_5$ & $-29.3854$ & $5.1$ \\
    \hline 
    $\hat{S}_6$ & $-16.1204$ & $116.4$ \\
    \hline 
    $\hat{S}_7$ & $-17.6195$ & $102.0$ \\
    \hline 
    $\hat{S}_8$ & $-22.2261$ & $70.1$ \\
    \Xhline{1pt}
    \end{tabular}
\end{center}
\end{table}

\subsection{Simulation Validation}\label{sec_SimVali}

\begin{figure*}[t]
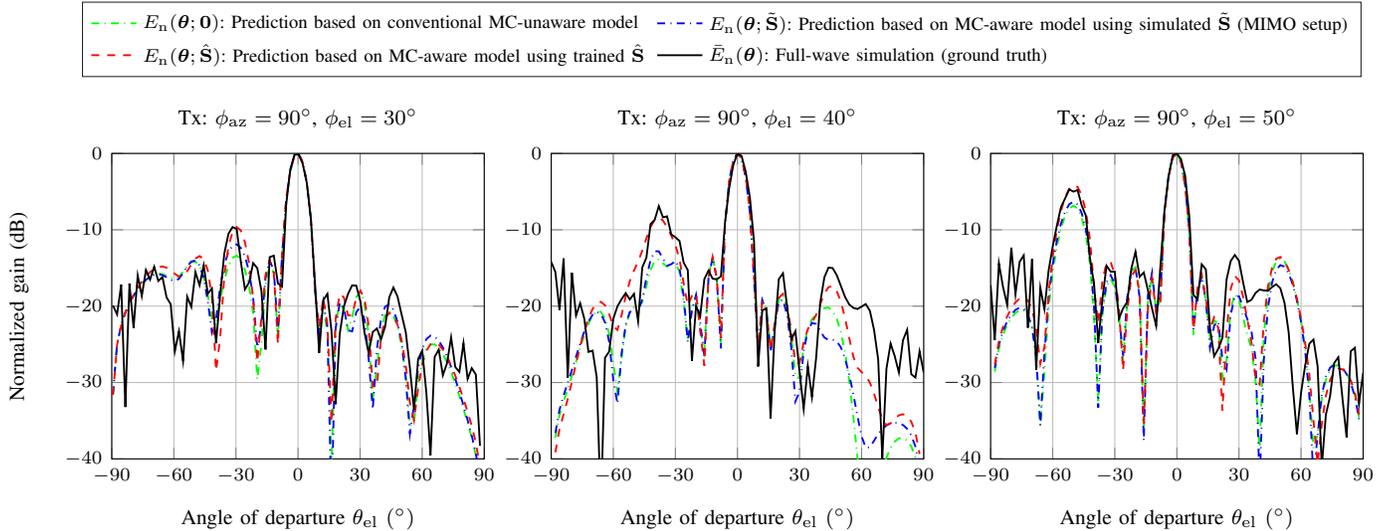

\centering
  \begin{minipage}[b]{1\linewidth}
      \include{figures/fig_simvali.tex}
      \vspace{-2.5em}
  \end{minipage}
  \caption{
	The comparison of the RIS radiation pattern obtained through the mutual coupling-unaware model, the mutual coupling-aware model using simulated $\tilde{\Sm}$ (MIMO setup), the mutual coupling-aware model using trained $\hat{\Sm}$, and the full-wave simulation. 
  }
  \label{fig_simvali}
\end{figure*}

Since we have learned a scattering matrix $\hat{\Sm}$ based on the simulated RIS radiation pattern in Fig.~\ref{fig_training}, we now evaluate the accuracy of the estimated $\hat{\Sm}$ as well as model~\eqref{eq_simpH2}. We conduct another 3 different full-wave simulations of RIS radiation pattern by varying the system setup. Specifically, we change the incident elevation angle of the Tx horn to $\phi_\mathrm{el}=30^\circ$, $40^\circ$, and $50^\circ$, respectively.\footnote{ {Note that in all full-wave simulations, we obtain the far-field radiation pattern reflected from the RIS, which characterized by~\acp{aod} only. Concurrently, the~\ac{em} transmission from the Tx to the RIS is determined based on the Tx's actual position, i.e., employing a near-field model. This aligns with the theoretical calculations outlined in~\eqref{eq_Econv}--\eqref{eq_hAoD} and ensures the consistency between the model and observations.}} The corresponding RIS beamforming is also changing according to the method in~\cite{Wang2023Wideband}.
Fig.~\ref{fig_simvali} compares the different theoretical predictions and the full-wave simulation (serving as ground truth) of RIS radiation patterns over these 3 different setups.
The evaluated theoretical results include the prediction based on the conventional mutual coupling-unaware model (i.e., ${E}_\mathrm{n}(\thetav;\mathbf{0})$), the mutual coupling-aware model using simulated $\tilde{\Sm}$ in the MIMO setup (i.e., ${E}_\mathrm{n}(\thetav;\tilde{\Sm})$), and the mutual coupling-aware model using trained $\hat{\Sm}$ (i.e., ${E}_\mathrm{n}(\thetav;\hat{\Sm})$).

Typically, for the normalized radiation pattern of an array, the predicting accuracy of its sidelobe (especially the strongest sidelobe) is a major concern. As demonstrated in Fig.~\ref{fig_simvali}, the prediction based on the mutual coupling-aware model using trained $\hat{\Sm}$ (red dashed curve) outperforms the other two theoretical predictions, showing the highest accuracy aligned with the full-wave simulations across the three tested setups.  {The corresponding prediction errors are listed in Table~\ref{tab3}, which are computed as
\begin{equation}\label{eq:PE1}
	\Pt\Et = \frac{\|\rv(\Qm)-\bar{\rv}\|_2}{\|\bar{\rv}\|_2},
\end{equation}
where~$\bar{\rv}\overset{\Delta}{=}[\bar{E}_\mathrm{n}(\thetav_1),\dots,\bar{E}_\mathrm{n}(\thetav_{L_\Rt})]^\TT\in\mathbb{R}^{L_\Rt}$,~$\rv(\Qm)\overset{\Delta}{=}[{E}_\mathrm{n}(\thetav_1;\Qm),\dots,{E}_\mathrm{n}(\thetav_{L_\Rt};\Qm)]^\TT\in\mathbb{R}^{L_\Rt}$, with~$L_\Rt$ denoting the sample size of these testing radiation patterns and~$\Qm\in\{\mathbf{0},\tilde{\Sm},\hat{\Sm}\}$. Here, we set~$L_\Rt=90$ by uniformly sampling over~$-90^\circ<\theta_\mathrm{el}<90^\circ$ with a~$2^\circ$ step size.} It is observed that the trained model~$\hat{\Sm}$ surpasses the other two models in predicting the RIS radiation pattern with higher accuracy. Particularly, at~$\phi_\mathrm{el}=40^\circ$, there is a reduction of ~\unit[10.7]{\%} in prediction error observed.
These results reveal that: (i) \textbf{the mutual coupling-aware communication model~\eqref{eq_simpH2} can correctly describe the mutual coupling effect in RIS}, thus presenting a more precise performance in predicting RIS radiation pattern compared to the mutual coupling-unaware model~$\Qm=\mathbf{0}$. (ii) \textbf{the proposed model training method can accurately estimate the scattering matrix~$\Sm$}, as the prediction result using trained~$\hat{\Sm}$ significantly and consistently outperforms that using simulated~$\tilde{\Sm}$ in MIMO setup.  {An additional observation is that at certain AoAs, such as $\phi_\mathrm{el}=40^\circ$, the deviation between the conventional model and the ground truth is much more significant than in the other two cases. This indicates an angle-selective property of the impact of mutual coupling.}

\begin{table}[t]
  \renewcommand{\arraystretch}{1.3}
  \begin{center}
  \caption{ {Prediction Error~$\mathrm{PE}$ of the Radiation Patterns in Fig.~\ref{fig_simvali}}}\vspace{-0.5em}
  \label{tab3}
  \begin{tabular}{  c !{\vrule width1pt} c !{\vrule} c !{\vrule} c }
    \Xhline{1pt}
    Incident Angle & $\phi_\mathrm{el}=30^\circ$ & $\phi_\mathrm{el}=40^\circ$ &  $\phi_\mathrm{el}=50^\circ$ \\
    \Xhline{1pt}
    $\Qm=\mathbf{0}$ & \unit[25.5]{\%} & \unit[44.4]{\%} & \unit[32.8]{\%} \\
    \hline 
    $\Qm=\tilde{\Sm}$ & \unit[26.0]{\%} & \unit[39.8]{\%} & \unit[33.6]{\%} \\
    \hline 
    $\Qm=\hat{\Sm}$ & \unit[23.8]{\%} & \unit[33.7]{\%} & \unit[27.9]{\%} \\
    \Xhline{1pt}
    \end{tabular}
    \begin{tablenotes}
     \item{\footnotesize \quad $\bullet$ The calculation of prediction errors is given by~\eqref{eq:PE1}.} 
   \end{tablenotes}
\end{center}
\end{table}

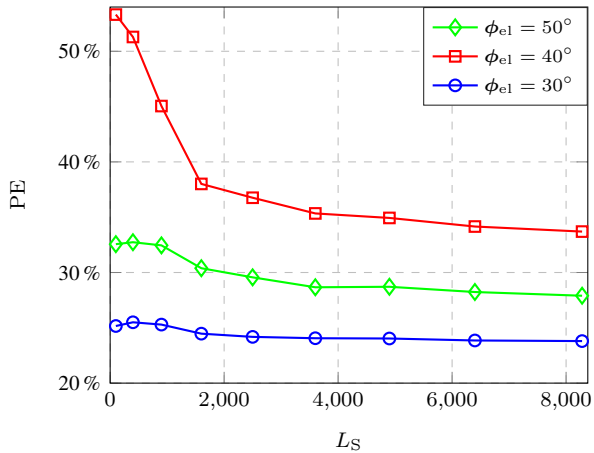
\begin{figure}[t]
    \centering
%
%
\begin{tikzpicture}

\begin{axis}[%
width=2.5in,
height=1.97in,
at={(0in,0in)},
scale only axis,
xmin=0,
xmax=8381,
xticklabel style = {font=\color{white!15!black},font=\footnotesize},
xlabel style={font=\color{white!15!black},font=\footnotesize},
xlabel={$L_{\St}$},
ymin=20,
ymax=54,
xmajorgrids,
xminorgrids,
ymajorgrids,
yminorgrids,
grid style={dashed},
yticklabel style = {font=\color{white!15!black},font=\footnotesize},
ylabel style={font=\color{white!15!black},font=\footnotesize},
ylabel={$\mathrm{PE}$},
ytick = {20,30,40,50},
yticklabels = {\unit[20]{\%},\unit[30]{\%},\unit[40]{\%},\unit[50]{\%}},
axis background/.style={fill=white},
legend style={at={(1,1)}, anchor=north east, legend cell align=left, align=left, draw=white!15!black, font=\scriptsize}
]
\addplot [color=green, mark=diamond, line width=0.8pt, mark options={solid, green}, mark size=3pt]
  table[row sep=crcr]{%
100	32.5621912715319\\
400	32.7374799946184\\
900	32.4525786465037\\
1600	30.3873734548523\\
2500	29.5626466892022\\
3600	28.6650893067578\\
4900	28.7090439699825\\
6400	28.2358262019104\\
8281	27.9\\
};
\addlegendentry{$\bm{\phi}_{\mathrm{el}}=50^\circ$}

\addplot [color=red, mark=square, line width=0.8pt, mark options={solid, red}]
  table[row sep=crcr]{%
100	53.3170978530637\\
400	51.2925777451676\\
900	45.0481338919971\\
1600	38.0088011750015\\
2500	36.7588668985091\\
3600	35.3404690207075\\
4900	34.9330244727582\\
6400	34.1566768103862\\
8281	33.7\\
};
\addlegendentry{$\bm{\phi}_{\mathrm{el}}=40^\circ$}

\addplot [color=blue, mark=o, line width=0.8pt, mark options={solid, blue}, mark size=2.2pt]
  table[row sep=crcr]{%
100	25.1635765762264\\
400	25.4993678749468\\
900	25.2903290518273\\
1600	24.470843773672\\
2500	24.1804907319502\\
3600	24.0579677466603\\
4900	24.0333206969724\\
6400	23.8551421434902\\
8281	23.8\\
};
\addlegendentry{$\bm{\phi}_{\mathrm{el}}=30^\circ$}

\end{axis}

\end{tikzpicture}%
    \vspace{-3em}
    \caption{ {Evaluation of the prediction error~$\mathrm{PE}$ (defined in~\eqref{eq:PE1}) versus the number of samples~$L_\St$ on simulated radiation pattern that used to estimate~$\hat{\Sm}$.}}
    \label{fig_samNum}
\end{figure}

 {
\subsection{Complexity Analysis}

\subsubsection{Resources Consumption of the Full-Wave Simulation}

\begin{table}[t]
  \renewcommand{\arraystretch}{1.3}
  \begin{center}
  \caption{ {Consumed Computational Resources of Full-Wave Simulation over Various RIS Sizes}}\vspace{-0.5em}
  \label{taba1}
  \begin{tabular}{  p{1.7cm}<{\centering} !{\vrule width1pt} c !{\vrule} c !{\vrule} c !{\vrule} c  }
    \Xhline{1pt}
    RIS Size & $5\times 5$ & $10\times 10$ &  $15\times 15$ & $20\times 20$ \\
    \Xhline{1pt}
    Total memory & \unit[7.4]{GB} & \unit[29.8]{GB} & \unit[63.9]{GB} & \unit[115]{GB}\\
    \hline 
    Elapsed time & \unit[8]{m} \unit[30]{s} & \unit[40]{m} \unit[51]{s} & \unit[2]{h} \unit[26]{m} \unit[53]{s} & \unit[6]{h} \unit[1]{m} \unit[17]{s}\\
    \hline 
    No. tetrahedra & 243124 & 906070 & 2098028 & 3624116 \\
    \hline 
    Matrix size & 1820821 & 6541935 & 15121957 & 26101753 \\
    \Xhline{1pt}
    \end{tabular}
\end{center}
\end{table}

Our proposed S-parameters estimation method is based on the observations obtained through the full-wave simulation. To evaluate the complexity of this full-wave simulation process, Table~\ref{taba1} summarizes the computational resources consumptions when simulating over various RIS sizes. Here, the processor of the workstation is Intel(R) Xeon(R) Gold 6230 CPU @\unit[2.10]{GHz} (2 processors) with installed RAM of \unit[256]{GB}, and the utilized simulation software is ANSYS HFSS Version 2021.1.0. It can be observed that as the RIS array becomes larger, the consumed computational resources significantly increase, making full-wave simulation on large-scale RISs impossible. However, since our scattering model is parameterized into limited amount of S-parameters, it is always feasible to perform full-wave simulations on a smaller RIS as long as all the defined S-parameters (see Fig.~\ref{fig_S028}) are involved. Nevertheless, this reduction can diminish model robustness, thus indicating a trade-off between the computational complexity and the simulation precision.

\subsubsection{Computational Complexity of the Model Training}
Now we evaluate the computational complexity of the proposed model training process. According to the gradient descent principle and the derivative calculation presented in Appendix, we can determine the computational complexity of the proposed scattering parameters estimation method as~$\mathcal{O}(L_\mathrm{S} N^2)$. The complexity of the estimation process increases quadratically with the RIS size and linearly with the number of radiation pattern observation samples. While the size of the RIS is fixed following the previous full-wave simulation step, the number of observation samples is adjustable, posing a potential impact on estimation accuracy. Fig.~\ref{fig_samNum} shows the prediction errors evaluated according to~\eqref{eq:PE1} with~$\Qm=\hat{\Sm}$ versus sample sizes~$L_\mathrm{S}$. It is observed that a greater~$L_\mathrm{S}$ can generally result in a lower prediction error, which also reveals a trade-off between the computational complexity and the estimation accuracy.
}

\section{Experimental Results}\label{sec_ER}

\begin{figure}[t]
  \centering
  \includegraphics[width=0.99\linewidth]{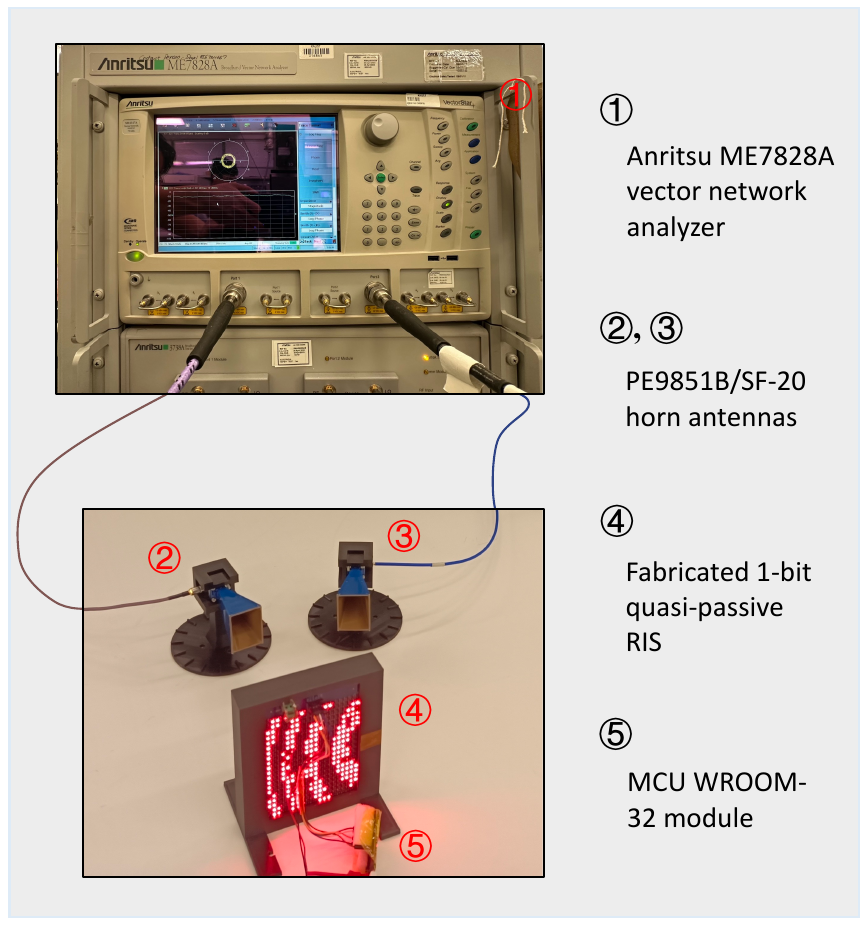}
  \caption{ 
     Experimental measurement setup for the fabricated RIS prototype. The S-parameter between two horns is measured using a vector network analyzer.
    }
  \label{fig_ES}
\end{figure}

Although the accuracy of the mutual coupling-aware model and the trained scattering matrix have been verified by full-wave simulation in Section~\ref{sec_SimVali}, further validations are performed in this section through experimental measurement on the fabricated RIS prototype described in Section~\ref{sec_RIS}.
The measurement setup is demonstrated in Fig.~\ref{fig_ES}. Here, we use two PE9851B/SF-20 horn antennas to transmit and receive wireless signals, while the fabricated 1-bit quasi-passive RIS is utilized to reflect the \ac{em} waves.
As mentioned, an MCU WROOM-32 module is used to control the beamforming at the RIS, which can be visualized through the LED array at the back surface of the RIS.
Based on this setup, the S-parameter between the Tx and Rx horns can be measured using an Anritsu ME7828A vector network analyzer. This real measured Tx-Rx S-parameter is denoted as~$\bar{S}_\mathrm{RT}$. Both the distance between Tx and RIS and the distance between Rx and RIS are set as $\unit[0.18]{m}$,  {which are within Rayleigh distance (approximately~\unit[2.14]{m}) of the fabricated RIS and form a near-field scenario}.

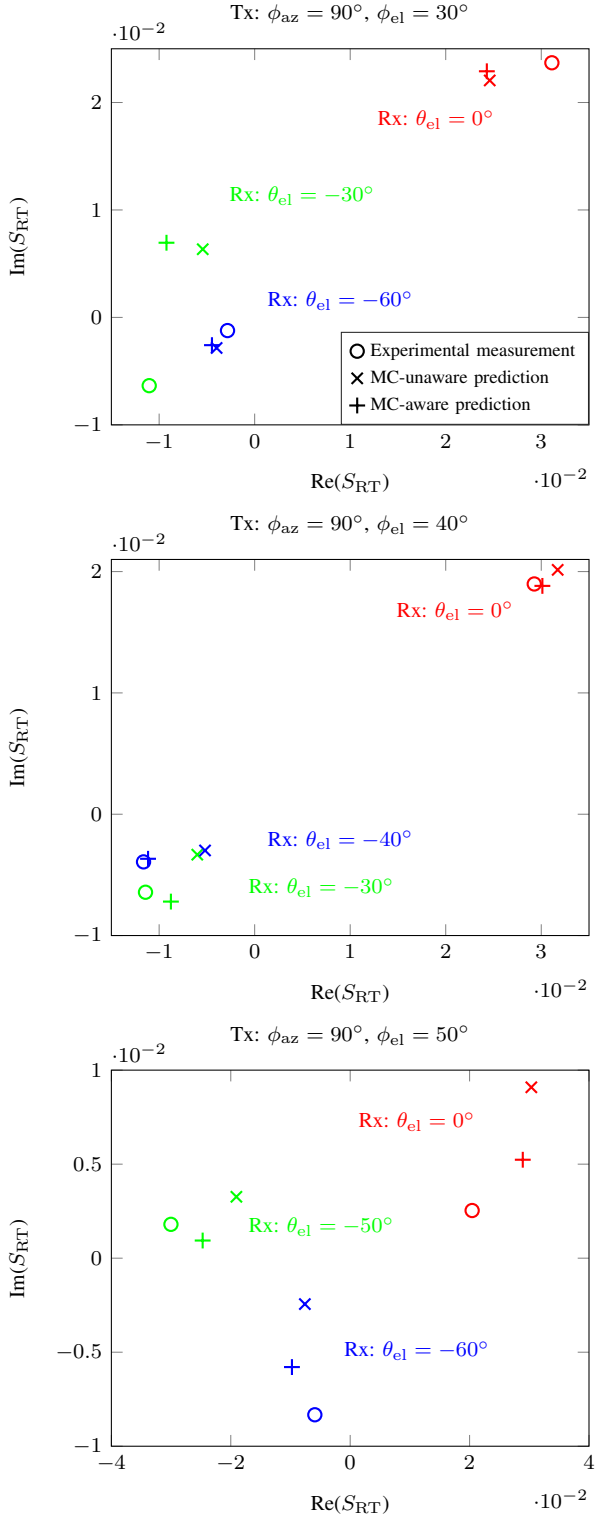
\begin{figure}[t]
\centering
  \begin{minipage}[b]{1\linewidth}
%
%
\begin{tikzpicture}

\begin{axis}[%
width=2.5in,
height=1.97in,
at={(0in,5.35in)},
scale only axis,
xmin=-0.015,
xmax=0.035,
xlabel style={font=\color{white!15!black},font=\footnotesize},
xticklabel style = {font=\color{white!15!black},font=\footnotesize},
xlabel={Re($S_\mathrm{RT}$)},
ymin=-0.01,
ymax=0.025,
ylabel style={font=\color{white!15!black},font=\footnotesize},
yticklabel style = {font=\color{white!15!black},font=\footnotesize},
ylabel={Im($S_\mathrm{RT}$)},
axis background/.style={fill=white},
title={{\footnotesize Tx: $\phi_\mathrm{az}=90^\circ$, $\phi_\mathrm{el}=30^\circ$}},
legend style={at={(1,0)}, anchor=south east, legend columns=1, legend cell align=left, align=left, draw=white!15!black, font=\scriptsize}
]
\addplot[only marks, mark=o, mark options={line width=0.8pt}, mark size=2.500pt, draw=black] table[row sep=crcr]{%
x	y\\
0.311120088102927	0.0236972323652668\\
};
\addlegendentry{Experimental measurement}

\addplot[only marks, mark=x, mark options={line width=0.8pt}, mark size=3.000pt, draw=black] table[row sep=crcr]{%
x	y\\
0.246064098661256	0.0220531009282365\\
};
\addlegendentry{MC-unaware prediction}

\addplot[only marks, mark=+, mark options={line width=0.8pt}, mark size=3.000pt, draw=black] table[row sep=crcr]{%
x	y\\
0.243034637977436	0.0229055659365641\\
};
\addlegendentry{MC-aware prediction}


\addplot[only marks, mark=o, mark options={line width=0.8pt}, mark size=2.5000pt, draw=red, forget plot] table[row sep=crcr]{%
x	y\\
0.0311120088102927	0.0236972323652668\\
};

\addplot[only marks, mark=x, mark options={line width=0.8pt}, mark size=3.000pt, draw=red, forget plot] table[row sep=crcr]{%
x	y\\
0.0246064098661256	0.0220531009282365\\
};

\addplot[only marks, mark=+, mark options={line width=0.8pt}, mark size=3.000pt, draw=red, forget plot] table[row sep=crcr]{%
x	y\\
0.0243034637977436	0.0229055659365641\\
};

\addplot[only marks, mark=o, mark options={line width=0.8pt}, mark size=2.5000pt, draw=green, forget plot] table[row sep=crcr]{%
x	y\\
-0.0110389780360829	-0.00635189910779714\\
};
\addplot[only marks, mark=x, mark options={line width=0.8pt}, mark size=3.000pt, draw=green, forget plot] table[row sep=crcr]{%
x	y\\
-0.00543497804111123	0.00634498976905336\\
};
\addplot[only marks, mark=+, mark options={line width=0.8pt}, mark size=3.000pt, draw=green, forget plot] table[row sep=crcr]{%
x	y\\
-0.00923052736446329	0.00694880518808808\\
};
\addplot[only marks, mark=o, mark options={line width=0.8pt}, mark size=2.5000pt, draw=blue, forget plot] table[row sep=crcr]{%
x	y\\
-0.00284229810610141	-0.00122538683509349\\
};
\addplot[only marks, mark=x, mark options={line width=0.8pt}, mark size=3.000pt, draw=blue, forget plot] table[row sep=crcr]{%
x	y\\
-0.00398359932321756	-0.0028432472445294\\
};
\addplot[only marks, mark=+, mark options={line width=0.8pt}, mark size=3.000pt, draw=blue, forget plot] table[row sep=crcr]{%
x	y\\
-0.00447045460753656	-0.00258182614635479\\
};
\node at (1.7in,1.6in){\color{red}\footnotesize{Rx: $\theta_\mathrm{el}=0^\circ$}};
\node at (1in,1.2in){\color{green}\footnotesize{Rx: $\theta_\mathrm{el}=-30^\circ$}};
\node at (1.2in,0.65in){\color{blue}\footnotesize{Rx: $\theta_\mathrm{el}=-60^\circ$}};
\end{axis}

\begin{axis}[%
width=2.5in,
height=1.97in,
at={(0in,2.675in)},
scale only axis,
xmin=-0.015,
xmax=0.035,
xlabel style={font=\color{white!15!black},font=\footnotesize},
xticklabel style = {font=\color{white!15!black},font=\footnotesize},
xlabel={Re($S_\mathrm{RT}$)},
ymin=-0.01,
ymax=0.021,
ylabel style={font=\color{white!15!black},font=\footnotesize},
yticklabel style = {font=\color{white!15!black},font=\footnotesize},
ylabel={Im($S_\mathrm{RT}$)},
axis background/.style={fill=white},
title={{\footnotesize Tx: $\phi_\mathrm{az}=90^\circ$, $\phi_\mathrm{el}=40^\circ$}},
]
\addplot[only marks, mark=o, mark options={line width=0.8pt}, mark size=2.5000pt, draw=red, forget plot] table[row sep=crcr]{%
x	y\\
0.0292684780986115	0.0189802025515883\\
};
\addplot[only marks, mark=x, mark options={line width=0.8pt}, mark size=3pt, draw=red, forget plot] table[row sep=crcr]{%
x	y\\
0.0317206147205877	0.0201432436229295\\
};
\addplot[only marks, mark=+, mark options={line width=0.8pt}, mark size=3pt, draw=red, forget plot] table[row sep=crcr]{%
x	y\\
0.030117684303436	0.0188207658687454\\
};
\addplot[only marks, mark=o, mark options={line width=0.8pt}, mark size=2.5000pt, draw=green, forget plot] table[row sep=crcr]{%
x	y\\
-0.0114271123285461	-0.00643067745205061\\
};
\addplot[only marks, mark=x, mark options={line width=0.8pt}, mark size=3pt, draw=green, forget plot] table[row sep=crcr]{%
x	y\\
-0.00600680574685077	-0.00333322651731481\\
};
\addplot[only marks, mark=+, mark options={line width=0.8pt}, mark size=3pt, draw=green, forget plot] table[row sep=crcr]{%
x	y\\
-0.00876754577622958	-0.00720142727330482\\
};
\addplot[only marks, mark=o, mark options={line width=0.8pt}, mark size=2.5000pt, draw=blue, forget plot] table[row sep=crcr]{%
x	y\\
-0.0116325365805848	-0.00392732929099488\\
};
\addplot[only marks, mark=x, mark options={line width=0.8pt}, mark size=3pt, draw=blue, forget plot] table[row sep=crcr]{%
x	y\\
-0.0051954478075328	-0.00298688036876889\\
};
\addplot[only marks, mark=+, mark options={line width=0.8pt}, mark size=3pt, draw=blue, forget plot] table[row sep=crcr]{%
x	y\\
-0.0111598351667579	-0.00367287381669273\\
};
\node at (1.8in,1.7in){\color{red}\footnotesize{Rx: $\theta_\mathrm{el}=0^\circ$}};
\node at (1.1in,0.25in){\color{green}\footnotesize{Rx: $\theta_\mathrm{el}=-30^\circ$}};
\node at (1.2in,0.5in){\color{blue}\footnotesize{Rx: $\theta_\mathrm{el}=-40^\circ$}};
\end{axis}

\begin{axis}[%
width=2.5in,
height=1.97in,
at={(0in,0in)},
scale only axis,
xmin=-0.04,
xmax=0.04,
xlabel style={font=\color{white!15!black},font=\footnotesize},
xticklabel style = {font=\color{white!15!black},font=\footnotesize},
xlabel={Re($S_\mathrm{RT}$)},
ymin=-0.01,
ymax=0.01,
ylabel style={font=\color{white!15!black},font=\footnotesize},
yticklabel style = {font=\color{white!15!black},font=\footnotesize},
ylabel={Im($S_\mathrm{RT}$)},
axis background/.style={fill=white},
title={{\footnotesize Tx: $\phi_\mathrm{az}=90^\circ$, $\phi_\mathrm{el}=50^\circ$}},
]
\addplot[only marks, mark=o, mark options={line width=0.8pt}, mark size=2.5000pt, draw=red, forget plot] table[row sep=crcr]{%
x	y\\
0.0204270523621157	0.00253443722044954\\
};
\addplot[only marks, mark=x, mark options={line width=0.8pt}, mark size=3pt, draw=red, forget plot] table[row sep=crcr]{%
x	y\\
0.030341242693599	0.00908778571504749\\
};
\addplot[only marks, mark=+, mark options={line width=0.8pt}, mark size=3pt, draw=red, forget plot] table[row sep=crcr]{%
x	y\\
0.0289096354851332	0.00523503853333625\\
};
\addplot[only marks, mark=o, mark options={line width=0.8pt}, mark size=2.5000pt, draw=green, forget plot] table[row sep=crcr]{%
x	y\\
-0.0300235663348732	0.00179955738161959\\
};
\addplot[only marks, mark=x, mark options={line width=0.8pt}, mark size=3pt, draw=green, forget plot] table[row sep=crcr]{%
x	y\\
-0.019059562107791	0.00326481974436513\\
};
\addplot[only marks, mark=+, mark options={line width=0.8pt}, mark size=3pt, draw=green, forget plot] table[row sep=crcr]{%
x	y\\
-0.0247096652886303	0.000938778185720541\\
};
\addplot[only marks, mark=o, mark options={line width=0.8pt}, mark size=2.5000pt, draw=blue, forget plot] table[row sep=crcr]{%
x	y\\
-0.00592153653583718	-0.00832730085712411\\
};
\addplot[only marks, mark=x, mark options={line width=0.8pt}, mark size=3pt, draw=blue, forget plot] table[row sep=crcr]{%
x	y\\
-0.00759013485550873	-0.00244157250470404\\
};
\addplot[only marks, mark=+, mark options={line width=0.8pt}, mark size=3pt, draw=blue, forget plot] table[row sep=crcr]{%
x	y\\
-0.00975423473920646	-0.00579046205401368\\
};
\node at (1.6in,1.7in){\color{red}\footnotesize{Rx: $\theta_\mathrm{el}=0^\circ$}};
\node at (1.1in,1.15in){\color{green}\footnotesize{Rx: $\theta_\mathrm{el}=-50^\circ$}};
\node at (1.6in,0.5in){\color{blue}\footnotesize{Rx: $\theta_\mathrm{el}=-60^\circ$}};
\end{axis}

\end{tikzpicture}%
      \vspace{-3em}
  \end{minipage}
  \caption{
	Comparison of the $S_{\mathrm{RT}}$ between Tx and Rx horns obtained by the experiment measurement, MC-unaware prediction, and MC-aware prediction.
  }
  \label{fig_expeComp}
\end{figure}

On the other hand, both communication model~\eqref{eq_Hconven} and~\eqref{eq_simpH2} (with trained $\hat{\Sm}$) can be adopted to predict this S-parameter~$S_\mathrm{RT}$. By comparing both theoretical predictions with the measured results, we can assess the precision of model~\eqref{eq_Hconven} and~\eqref{eq_simpH2}. Specifically, the S-parameter between Tx and Rx horns can be theoretically calculated as
\begin{align}
	\text{\textbf{MC-unaware:} } & S_\mathrm{RT}^\mathrm{conv} = \hv_\mathrm{RI}\Thetam\hv_\mathrm{IT}, \\
	\text{\textbf{MC-aware:} } & S_\mathrm{RT}^\mathrm{new} = \hv_\mathrm{RI}\big(\Thetam^{-1}-\hat{\Sm}\big)^{-1}\hv_\mathrm{IT}.
\end{align}
Here, each entry of the wireless channels~$\hv_\mathrm{IT}\in\mathbb{C}^{N\times 1}$ and~$\hv_\mathrm{RI}\in\mathbb{C}^{N\times 1}$ are computed as
\begin{align}
	h_{\mathrm{IT},n} &= \frac{\cos^{q_\mathrm{f}}(\theta_{\mathrm{T},n})}{\|\pv_n-\pv_\mathrm{T}\|_2}e^{-jk_0\|\pv_n-\pv_\mathrm{T}\|_2},\\
	h_{\mathrm{RI},n} &= \frac{\cos^{q_\mathrm{e}}(\theta_{\mathrm{e},n}) \cos^{q_\mathrm{f}}(\theta_{\mathrm{R},n})}{\|\pv_n-\pv_\mathrm{R}\|_2}e^{-jk_0\|\pv_n-\pv_\mathrm{R}\|_2},
\end{align}
where $\pv_\mathrm{T}$ and~$\pv_\mathrm{R}$ stand for the positions of the Tx and Rx horns, respectively.  {Note that these channels~$\hv_\mathrm{IT}$ and~$\hv_\mathrm{RI}$ have been adjusted to the near-field channel model to fit the measurement setup.}
Before the final comparison, we process the following minimization to remove the global phase shift and amplitude loss:
\begin{equation}\label{eq_Glo}
	\{\hat{\alpha},\hat{\beta}\} = \arg\min_{\alpha,\beta}\ \sum_{i=1}^{D}|\alpha e^{j\beta} S_{\mathrm{RT},i}^\mathrm{X}-\bar{S}_{\mathrm{RT},i}|^2,
\end{equation}
where~$i$ indices the prediction or measurement over different system setups and~$\mathrm{X}\in\{\mathrm{conv},\mathrm{new}\}$. 
Finally, we adopt $\hat{S}_{\mathrm{RT},i}^\mathrm{X}=\hat{\alpha} e^{j\hat{\beta}} S_{\mathrm{RT},i}^\mathrm{X},\ i=1,2,\dots,$ as the theoretical prediction for both models.

\begin{table}[t]
  \renewcommand{\arraystretch}{1.3}
  \begin{center}
  \caption{ {Prediction Error of the Tx-Rx S-parameters in Fig.~\ref{fig_expeComp}}}\vspace{-1em}
  \label{tab4}
  \begin{tabular}{  c !{\vrule width1pt} c !{\vrule} c !{\vrule} c }
    \Xhline{1pt}
    Incident Angle & $\phi_\mathrm{el}=30^\circ$ & $\phi_\mathrm{el}=40^\circ$ &  $\phi_\mathrm{el}=50^\circ$ \\
    \Xhline{1pt}
    $\Big(\sum\limits_{i=1}^3|\hat{S}_{\mathrm{RT},i}^\mathrm{conv}-\bar{S}_{\mathrm{RT},i}|^2\Big)^{{1}/{2}}$ & 0.0155 & 0.0094 & 0.0173 \\
    \hline 
    $\Big(\sum\limits_{i=1}^3|\hat{S}_{\mathrm{RT},i}^\mathrm{new}-\bar{S}_{\mathrm{RT},i}|^2\Big)^{{1}/{2}}$ & 0.0153 & 0.0035 & 0.0119 \\
    \Xhline{1pt}
    \end{tabular}
\end{center}
\end{table}

The measured and predicted S-parameters over different setups are plotted in Fig.~\ref{fig_expeComp}  {and the corresponding numerical errors between the measurements and predictions are computed and presented in Table~\ref{tab4}.} 
Here, we show the results of 3 different setups of the Tx horn, which correspond to the setups in Fig.~\ref{fig_simvali}. 
For each Tx setup, we test the cases where the Rx horn is placed at the 3 chosen \acp{aod} $\theta_\mathrm{el}$ corresponding to the peak values in Fig.~\ref{fig_simvali}, aiming to attain higher SNRs. 
 {As shown in Table~\ref{tab4}, the mutual coupling-aware model consistently offers more accurate predictions than the conventional model when benchmarked against real measured data. Interestingly, for $\phi_\mathrm{el}=30^\circ$, the mutual coupling-aware model yields only minor improvements over the conventional model. This can be attributed to the minimal impact of mutual coupling in this configuration, where the conventional model already provides accurate predictions of RIS radiation, as evidenced in Table~\ref{tab3}. However, for $\phi_\mathrm{el}=40^\circ$ and $\phi_\mathrm{el}=50^\circ$, the mutual coupling-aware model exhibits significantly lower prediction errors.} This observation further \textbf{validates the correctness of the scattering matrix-based model~\eqref{eq_simpH2} and the accuracy of our estimation on $\Sm$}.


\section{Conclusion}

This paper conducts a realistic evaluation of the mutual coupling in RIS-aided communication based on an authentic RIS prototype. Adopting the scattering matrix-based model, we parameterize the RIS scattering matrix by analyzing the geometric configuration of RIS reflecting elements. A practical training approach for these scattering parameters is proposed leveraging a single 3D full-wave simulation of the RIS radiation pattern. Both full-wave simulations and experimental measurements are carried out to verify the accuracy of the trained model. The results of this work validate the correctness of the RIS-aided wireless communication model based on scattering matrices. Furthermore, it offers an effective method to estimate RIS scattering parameters, foreseeing substantial potential in applications such as RIS-aided channel estimation, beamforming, radio localization, etc.  {Nonetheless, the proposed model training method requires full-wave simulation, which is time-consuming and thus not suitable for online mutual coupling calibration. Future work can explore more complex RIS designs, such as multi-polarization, nonuniform array configurations, and more realistic modeling that considers the frequency and configuration dependencies of mutual coupling.}
 {
\appendix
This appendix demonstrates the convexity of~\eqref{eq:opt2}. The following derivations are based on the complex-valued matrix differentiation theory~\cite{Zhang2017Matrix}. Considering a complex-value function~$f(\xv)$:~$\mathbb{C}^{N_x}\rightarrow \mathbb{R}$, the adopted differentiation rule derives complex derivatives by treating~$\xv$ and~$\xv^*$ as two independent variables. In addition, for~$y\in\mathbb{C}$,~$\yv\in\mathbb{C}^{N_y}$, and~$\xv\in\mathbb{C}^{N_x}$, we use notations~$\frac{\partial y}{\partial \xv}\in\mathbb{C}^{N_x\times 1}$, $\frac{\partial \yv}{\partial \xv}\in\mathbb{C}^{N_y\times N_x}$ to denote the first-order partial derivatives.

We first define the unknown variables into the vector form as~$\sv\overset{\Delta}{=}[S_0,\dots,S_8]^\TT$ and~$\vv\overset{\Delta}{=}[\sv^\TT,\varepsilon]^\TT$. The domain of~$\vv$ is the manifold~$\mathbb{C}^9\times\mathbb{R}^+$, which is a convex set. Based on~\eqref{eq:f}, we define 
\begin{multline}
	g_\ell(\vv) = \Big(\hv_{\mathrm{out}}^\TT(\thetav_\ell)\big(\Thetam\!+\!\Thetam\big(\sum_{i=0}^8S_i\Am_i\big)\Thetam\big)\hv_\mathrm{in}\Big)\!\\ \Big(\hv_{\mathrm{out}}^\HH(\thetav_\ell)\big(\Thetam^*\!+\!\Thetam^*\big(\sum_{i=0}^8S_i^*\Am_i\big)\Thetam^*\big)\hv_\mathrm{in}^*\Big)\!-\!\varepsilon\bar{E}^2_\mathrm{n}(\bm{\theta}_\ell), \\
		\ell=1,\dots,L_\St.
\end{multline} 
Then we have
\begin{align}
	\frac{\partial g_\ell(\vv)}{\partial \sv} &\!=\! \av_\ell\Big(\hv_{\mathrm{out}}^\HH(\thetav_\ell)\big(\Thetam^*\!+\!\Thetam^*\big(\sum_{i=0}^8S_i^*\Am_i\big)\Thetam^*\big)\hv_\mathrm{in}^*\Big),\\
	\frac{\partial g_\ell(\vv)}{\partial \sv^*} &= \av_\ell^*\Big(\hv_{\mathrm{out}}^\TT(\thetav_\ell)\big(\Thetam\!+\!\Thetam\big(\sum_{i=0}^8S_i\Am_i\big)\Thetam\big)\hv_\mathrm{in}\Big),\\
	\frac{\partial g_\ell(\vv)}{\partial \varepsilon} &= \frac{\partial g_\ell(\vv)}{\partial \varepsilon^*} = \bar{E}^2_\mathrm{n}(\bm{\theta}_\ell),
\end{align}
where~$[\av_\ell]_i = \hv_{\mathrm{out}}^\TT(\thetav_\ell)(\Thetam\!+\!\Thetam\Am_i\Thetam)\hv_\mathrm{in}$. Therefore, we can derive
\begin{align}
	\frac{\partial^2 g_\ell(\vv)}{\partial \vv \partial \vv} &\!=\! \begin{bmatrix}
		\frac{\partial^2 g_\ell(\vv)}{\partial \sv\partial\sv} & \frac{\partial^2 g_\ell(\vv)}{\partial \varepsilon\partial\sv}\\[2pt]
		\big(\frac{\partial^2 g_\ell(\vv)}{\partial \sv\partial\varepsilon}\big)^\TT & \frac{\partial^2 g_\ell(\vv)}{\partial \varepsilon\partial\varepsilon}
	\end{bmatrix} \!= \mathbf{0}_{10\times 10},\\
	\frac{\partial^2 g_\ell(\vv)}{\partial \vv^* \partial \vv} &\!=\! \begin{bmatrix}
		\frac{\partial^2 g_\ell(\vv)}{\partial \sv^*\partial\sv} & \frac{\partial^2 g_\ell(\vv)}{\partial \varepsilon^*\partial\sv}\\[2pt]
		\big(\frac{\partial^2 g_\ell(\vv)}{\partial \sv^*\partial\varepsilon}\big)^\TT & \frac{\partial^2 g_\ell(\vv)}{\partial \varepsilon^*\partial\varepsilon}
	\end{bmatrix} \!=\! \begin{bmatrix}
		\av_\ell\av_\ell^\HH & \mathbf{0}_{9\times 1}\\[3pt]
		\mathbf{0}_{1\times 9} & 0
	\end{bmatrix},\\
	\frac{\partial^2 g_\ell(\vv)}{\partial \vv \partial \vv^*} &\!=\! \begin{bmatrix}
		\frac{\partial^2 g_\ell(\vv)}{\partial \sv\partial\sv^*} & \frac{\partial^2 g_\ell(\vv)}{\partial \varepsilon\partial\sv^*}\\[2pt]
		\big(\frac{\partial^2 g_\ell(\vv)}{\partial \sv\partial\varepsilon^*}\big)^\TT & \frac{\partial^2 g_\ell(\vv)}{\partial \varepsilon\partial\varepsilon^*}
	\end{bmatrix} \!=\! \begin{bmatrix}
		\av_\ell^*\av_\ell^\TT & \mathbf{0}_{9\times 1}\\[3pt]
		\mathbf{0}_{1\times 9} & 0
	\end{bmatrix},\\
	\frac{\partial^2 g_\ell(\vv)}{\partial \vv^* \partial \vv^*} &\!=\! \begin{bmatrix}
		\frac{\partial^2 g_\ell(\vv)}{\partial \sv^*\partial\sv^*} & \frac{\partial^2 g_\ell(\vv)}{\partial \varepsilon^*\partial\sv^*}\\[2pt]
		\big(\frac{\partial^2 g_\ell(\vv)}{\partial \sv^*\partial\varepsilon^*}\big)^\TT & \frac{\partial^2 g_\ell(\vv)}{\partial \varepsilon^*\partial\varepsilon^*}
	\end{bmatrix} \!= \mathbf{0}_{10\times 10}.
\end{align}
According to~\cite[Eq.~(3.5.10)]{Zhang2017Matrix}, the full complex Hessian matrix of~$g_\ell(\vv)$ is given by
\begin{align}
	\Hm_{g_\ell(\vv)} = \mathrm{blkdiag}(\av_\ell\av_\ell^\HH,0,\av_\ell^*\av_\ell^\TT,0) \succeq \mathbf{0}.
\end{align}
Thus,~$g_\ell(\vv)$ is convex,~$\forall\ell=1,\dots,L_\St$, based on the second-order condition of convexity~\cite{Boyd2004Convex}.

Now we consider~$f(\vv)$ in~\eqref{eq:f}, which can be written as~$f(\vv) = \|\gv(\vv)\|_2$ with~$\gv(\vv)\overset{\Delta}{=}[g_1(\vv),\dots,g_{L_\St}(\vv)]^\TT$. Since the~$\ell_2$-norm~$\|\cdot\|_2$ is convex and nondecreasing in each argument~$g_\ell(\vv)$, along with the fact that each~$g_\ell(\vv)$ is convex, the function~$f(\vv)$ is convex according to the vector composition rule~\cite[Eq.~(3.15)]{Boyd2004Convex}. As a conclusion,~\eqref{eq:opt2} is a convex optimization problem.
}
\bibliography{references}
\bibliographystyle{IEEEtran}

\end{document}